# Quantitative Analytical Model for Scattering-type Scanning Near-field Optical Spectroscopy

Kirill V. Voronin[1,2], Iker Herrero León[3], Rainer Hillenbrand[3,4], Alexey Y. Nikitin[1,4,†]

[1]Donostia International Physics Center (DIPC), Donostia-San Sebastián 20018, Spain
[2]Universidad del Pais Vasco/Euskal Herriko Unibertsitatea, 20080 Donostia/San Sebastián, Basque Country, Spain
[3]CIC nanoGUNE BRTA, Tolosa Hiribidea 76, 20018 Donostia-San Sebastian, Spain
[4]IKERBASQUE, Basque Foundation for Science, Bilbao 48013, Spain

[†]Contact author: alexey@dipc.org

**Abstract:** Scattering-type scanning near-field optical microscopy (s-SNOM) is a versatile technique in nanooptics, enabling local probing of optical responses beyond the diffraction limit from vis to THz frequencies. Its theoretical modeling based on tip-sample interactions typically relies on computationally intensive numerical methods or phenomenological models with empiric fitting parameters, complicating spectral analysis and interpretation. Developing a rigorous quantitative analytical model remains a significant challenge in near-field microscopy. Here, we introduce an accurate analytical solution for the prolate spheroid model of s-SNOM in the quasi-electrostatic limit. We validate our solution through comparisons with numerical simulations and experimental spectra. Due to its higher computational efficiency compared to numerical simulation and higher accuracy compared to phenomenological solutions, our solution for spheroid model facilitates spectrum prediction and interpretation for homogeneous bulk samples, enables systematic exploration of parameter effects, and supports data generation for machine learning applications. Furthermore, the generality of our approach allows straightforward extension to more complex nanostructures.

**Introduction.** Scattering-type scanning near-field optical microscopy (s-SNOM) is an advanced nano-optical imaging and spectroscopy technique that achieves spatial resolution far beyond the optical diffraction limit [1]. By employing a sharp atomic force microscope (AFM) tip illuminated by infrared or visible radiation, s-SNOM enables probing optical properties at the nanoscale through the near-field interaction [2, 3]. Since its early experimental demonstrations in 1990s [4-8], s-SNOM has evolved significantly allowing operating at cryogenic temperatures [9-11], external magnetic fields [12] and in liquid environment [13, 14] to explore solid state [15, 16], chemical [17, 18] and biological [19, 20] phenomena.

The development of a s-SNOM theory has been necessary for a comprehensive analysis and interpretation of near-field measurements. Due to the deeply subwavelength scale of the near-field tip-sample interaction, the quasi-electrostatic approximation (that neglects the retardation effects) can be reasonable, especially for infrared and THz frequencies, significantly reducing computational complexity [2, 6, 21]. Early theoretical approaches to interpreting s-SNOM data relied heavily on analytical models, with the simplest one being the point-dipole model (PDM) [21]. In this model, the tip is approximated by a small sphere with a radius of the order of the tip apex curvature. The tip-sample interaction is then represented by a single oscillating dipole interacting with image dipoles in the sample. This approach allows one to quickly qualitatively estimate an s-SNOM near-field signal [21, 22]. However, this simplicity sacrifices quantitative accuracy due to ignoring the finite tip size, higher-order multipole interactions, which are particularly relevant for materials with a strong resonance response, such as e.g., SiC or $SiO_2$ [23-26]. To address the finite size of a tip, the finite dipole model (FDM) was developed, which assumes that the electric field of the tip can be approximated by the field of several point charges placed at the tip axis [24]. The FDM allows one to obtain a quantitative fitting of the measured near-field spectra even for highly resonant samples, but it contains a phenomenologically introduced g-parameter denoting an effective part of the polarization charge participating in the near-field interaction that is not rigorously calculated and has to be determined empirically.

In contrast to analytical models, requiring empirical adjustments for quantitative accuracy, computational modellings based on, e.g. finite-difference time-domain (FDTD) and finite-element method (FEM) or boundary element method (BEM) [26, 27] potentially offer a rigorous quantitative description of an arbitrary tip and sample geometries [28-33]. However, these methods are computationally intensive due to the fine spatial discretization (particularly, around a tip apex) of the computational domains and require extensive computation time. Hybrid, semi-analytical models that simplify numerical simulations by selecting optimal coordinates and preliminary analytical derivations [34, 35] achieve a compromise between the speed of analytical calculations and the accuracy of numerical simulations, but also suffer from the shortcomings of both approaches.

Here, we present an exact analytical solution to the quasi-electrostatic prolate spheroid model, eliminating the need for phenomenological parameters and extensive numerical calculations. Our approach precisely determines the charge distribution on the surface of a spheroid with arbitrary dielectric permittivity placed in a uniform electric field above a sample surface with arbitrary dielectric permittivity. Previously, within the context of s-SNOM modeling, similar scenarios were addressed only numerically [34, 36, 37] or through approximate methods involving phenomenological coefficients [24]. While the electrostatic problem of a spheroid in an external field has been studied extensively, earlier analyses predominantly focused on special cases such as perfectly conducting spheroids subjected either to uniform fields or fields of point charges [38-40]. Furthermore, these approaches commonly relied on the image charge method, which was proved inadequate for charges situated near the spheroid surface [38]. Attempts to extend this method to scenarios involving close external charges resulted in significantly more complex calculations without fully resolving inherent divergences [40-42]. In contrast, our solution accurately computes the charge distribution at the spheroid surface with arbitrary dielectric properties near the sample, avoiding divergences associated with closely positioned charges.

Our solution provides accurate results that match FEM simulation of the system with the same geometry and materials parameters, while significantly reducing computation time by a factor of $10^3$ (from days to minutes). This substantial improvement enables a rapid analysis of how the key model parameters, including spheroid curvature radius and length, tapping amplitude, and dielectric permittivities of both the spheroid and the sample, influence the results. In particular, by conducting this analysis, we identify the radius of curvature as well as the minimum distance between the spheroid apex and the sample surface as the most critical model parameters. To illustrate the practical utility of our spheroid model, we compare the calculated spectra with experimental data for two representative materials, poly(methyl methacrylate) (PMMA) and quartz, which exhibit weak and strong Lorentz resonances, respectively.

**Results and discussions.** s-SNOM is based on AFM, wherein a vertically oscillating tip is illuminated by a focused beam of electromagnetic radiation (illustrated in Fig. 1a). The amplitude and phase of the elastically scattered light from the tip are measured interferometrically. The incident radiation can be either monochromatic or broadband. Spectroscopy of broadband scattered light with an asymmetric Fourier transform spectrometer is called nanoscale Fourier transform infrared (nano-FTIR) spectroscopy. Because the tip's polarizability is influenced both by the external electromagnetic field and by near-field interactions with the sample, the scattered signal encodes information about the sample's optical properties. To isolate the near-field contribution from the far-field background, the interferometric detector signal is demodulated at the $n^{\text{th}}$ (typically, second or higher) harmonic of the tip's oscillation frequency. These higher harmonics capture nonlinear variations in the induced charge distribution at the tip apex as the tip–sample distance is modulated by the tip oscillation. As a result, the apex geometry plays a dominant role in determining the near-field response, while the overall tip shape is of lesser importance. On the other hand, the exact geometry of the tip is not precisely known and may evolve during the measurement process due to mechanical deformation of the apex or chemical degradation, which can alter its effective permittivity. As a result, all s-SNOM models necessarily rely on idealized geometries and approximated material parameters to simulate the experimental conditions. To this end, the tip is commonly approximated as a prolate spheroid (see Fig. 1b). This approximation strikes a balance between physical rigor—being sufficiently accurate to reproduce experimental spectra—and computational efficiency [34, 36, 37, 43, 44]. Beyond its use in numerical simulations, the spheroidal geometry underpins the analytical FDM [24], which also captures key experimental features, albeit with the introduction of empirical parameters.

In typical s-SNOM configurations, both the tip curvature radius and the amplitude of the mechanical tip oscillation are much smaller than the illumination wavelength. Under these conditions, the quasi-electrostatic approximation, neglecting magnetic field contributions and retardation effects, is applicable to the prolate spheroid model. The core of the mathematical formulation involves computing the dipole moment, $\mathbf{p}$, of the spheroidal tip, induced by the incident electric field, $\mathbf{E}_{inc}$, from which the far-field scattered electric field, $\mathbf{E}_{sca}$, can be expressed as (see Supporting Information, Section 1) [24, 45]:

$$E_{sca}(H) \sim (1+\varsigma r_p) p_z(H) = (1+\varsigma r_p)^2 \alpha_z(H) E_{inc,z} \quad (1)$$

Here $\alpha_z(H)$ is the tip polarizability, the relation between the tip dipole moment and the external electric field; $r_p$ denotes the Fresnel reflection coefficient for p-polarized light; and $\varsigma$ is a reflected wave contribution modifier that is used to account for the influence of far-field-scale sample surface inhomogeneity, focused beam field inhomogeneity, and other far-field corrections. In this work, we always use $\varsigma = 1$. The prefactor $(1+\varsigma r_p)^2$ accounts for the reflection of both the incident and scattered fields at the sample interface. Note, here we consider only the dipole component along the spheroid's major axis, $p_z(H)$, as it dominates over transverse components due to the tip's elongated geometry. Also, we take into account only the vertical component of the incident electric field, as it makes the main contribution to the charge redistribution. Both $E_{sca}(H)$ and $\alpha_z(H)$ depend on the time-varying tip–sample distance, modeled as $H(t) = H_0 + A(1-\cos\Omega t)$, where $H_0$ is the minimal distance between tip apex and sample surface, $A$ is the tip oscillation amplitude, and $\Omega$ is the oscillation frequency. The detected near-field signal corresponds to the $n^{\text{th}}$ harmonic ($n = 2, 3$ or $4$) of the scattered field, extracted via Fourier decomposition (see Supporting Information, Section 1) [46]:

$$S_n \sim \Omega \int_0^{\frac{2\pi}{\Omega}} E_{sca}(H) e^{-in\Omega t} dt \quad (2)$$

Analogously, we define the $n^{\text{th}}$ harmonic of the spheroid polarizability, $\alpha_{nz}$:

$$\alpha_{nz} = \Omega \int_0^{\frac{2\pi}{\Omega}} \alpha_z(H) e^{-in\Omega t} dt \quad (3)$$

To eliminate proportionality constants, the near-field signal is normalized against that from a reference material with frequency-independent permittivity (at least within the spectral range of interest). This yields the normalized near-field signal [1]:

$$\sigma_n = \frac{S_{s,n}}{S_{r,n}} = \frac{(1+\varsigma r_{ps})^2 \alpha_{s,nz}}{(1+\varsigma r_{pr})^2 \alpha_{r,nz}} \quad (4)$$

where $S_{s,n}$, $r_{ps}$, $\alpha_{s,nz}$ refer to the signal complex amplitude, reflection coefficient and $n^{\text{th}}$ harmonic of the spheroid polarizability of the sample respectively and $S_{r,n}$, $r_{pr}$, $\alpha_{r,nz}$ correspond to that of the reference material.

To determine the main ingredient of the model – the polarizability of the spheroid – we calculate the induced charge distribution on the surface of the spheroid, placed near the sample surface in a uniform external electric field, as shown in Fig. 1b. To obtain the charge distribution on the spheroid surface, we solve Laplace's equation with boundary conditions ensuring continuity of the potential and the normal to the surface component of the electric displacement field, **D**, at both sample and tip surfaces. Calculations are simplified by employing spheroidal coordinates. Detailed derivations appear in Supporting Information, Sections 2, 3; the derivations are made using the mathematical relationships given in the following sources [47-51]. Here we present only the final expressions, which are sufficient to calculate the spheroid's polarizability z-component:

$$\alpha_z(H) = 2ac\vartheta_1(H) + \alpha_{z0}, \tag{5}$$

where $\alpha_{z0}$ is the term related to the polarization of spheroid in free space, the latter is independent on the interaction with the surface of the sample, and therefore, it does not contribute to the demodulated signal; $a = L/2$ is the spheroid major semi-axis, $L$ is the spheroid full length, $c = \sqrt{a^2 - b^2}$, $b$ is the spheroid minor semi-axis; and $\vartheta_n$ are determined by the solution of the following linear system of equations:

$$\sum_{n=1}^{\infty} \mathcal{M}_{mn}\,\vartheta_n = C_m \tag{6}$$

where the matrix, $\mathcal{M}_{mn}$, and the free term, $C_m$, read

$$\mathcal{M}_{mn} = \delta_{mn} - \frac{\beta(\varepsilon_T - 1)(2n+1)J_{nm}}{2P_m(\xi_0)\left(\varepsilon_T \frac{Q_m(\xi_0)}{P_m(\xi_0)} - \frac{\xi_0 Q_m(\xi_0) - Q_{m-1}(\xi_0)}{\xi_0 P_m(\xi_0) - P_{m-1}(\xi_0)}\right)} \tag{7}$$

$$C_m = -\frac{\beta c(\varepsilon_T - 1)^2 J_{1m}}{4P_m(\xi_0)\left(\varepsilon_T Q_1(\xi_0) - \xi_0 Q_0(\xi_0) + \frac{\xi_0^2}{\xi_0^2 - 1}\right)\left(\varepsilon_T \frac{Q_m(\xi_0)}{P_m(\xi_0)} - \frac{\xi_0 Q_m(\xi_0) - Q_{m-1}(\xi_0)}{\xi_0 P_m(\xi_0) - P_{m-1}(\xi_0)}\right)} \tag{8}$$

Note that despite Eq. (6) containing an infinite number of variables, only the first one is related to the dipole moment of the spheroid. In (7,8) $\varepsilon_T$ is the dielectric permittivity of the tip, $\beta = \frac{\varepsilon_s - 1}{\varepsilon_s + 1}$ is the electro-static Fresnel reflection coefficient, $\varepsilon_s$ is the dielectric permittivity of the sample, $P_n(x)$ and $Q_n(x)$ are Legendre functions of the first and second kind respectively, and $\xi_0 = \frac{a}{c}$ is the first coordinate of all point of the spheroid surface in the spheroidal coordinate system. In Eqs. (7,8) we introduce the following integral, which determines the link between the coordinates of the image charges and the coordinates of charges at the surface of the spheroid

$$J_{nm} = \int_{-1}^{1} d\eta P_n(\eta) P_m(\eta_q) Q_m(\xi_q) \tag{9}$$

where $(\xi, \eta)$ are the coordinates of the charges at the surface of the spheroid, and $(\xi_q, \eta_q)$ are the coordinates of the image charges at the coordinate system of the spheroid; $\eta_q = \frac{\psi}{\xi_q}$, $\psi = 2\frac{a+H}{c} - \xi_0\eta$, and $\xi_q$ reads

$$\xi_q = \sqrt{\frac{1 + \psi^2 + \chi^2 + \sqrt{(1 + \psi^2 + \chi^2)^2 - 4\psi^2}}{2}} \tag{10}$$

where $\chi = (\xi_0^2 - 1)(1 - \eta^2)$. When $m$ and $n$ tend to infinity, $\mathcal{M}_{mn}$ tends to a Kronecker symbol, $\delta_{mn}$, therefore, $\vartheta_n$ tends to $C_n$. On the other hand, when $n$ tends to infinity, $C_n$ tends to zero. Equations (5)-(10) provide an exact analytical solution of the spheroid problem. However, practical numerical evaluation requires performing integrals numerically and solving an infinite system of equations. To numerically address the infinite system in Eq. (6), a truncation approach can be applied by introducing a cutoff parameter, $N$, and considering only the first $N$ equations, assuming $\vartheta_n = 0$ for $n > N$. Importantly, our theoretical approach also holds for uniaxial samples with the axis perpendicular to the surface, except the expression for $\beta$, which should be modified as shown in Supporting Information, Section 2.

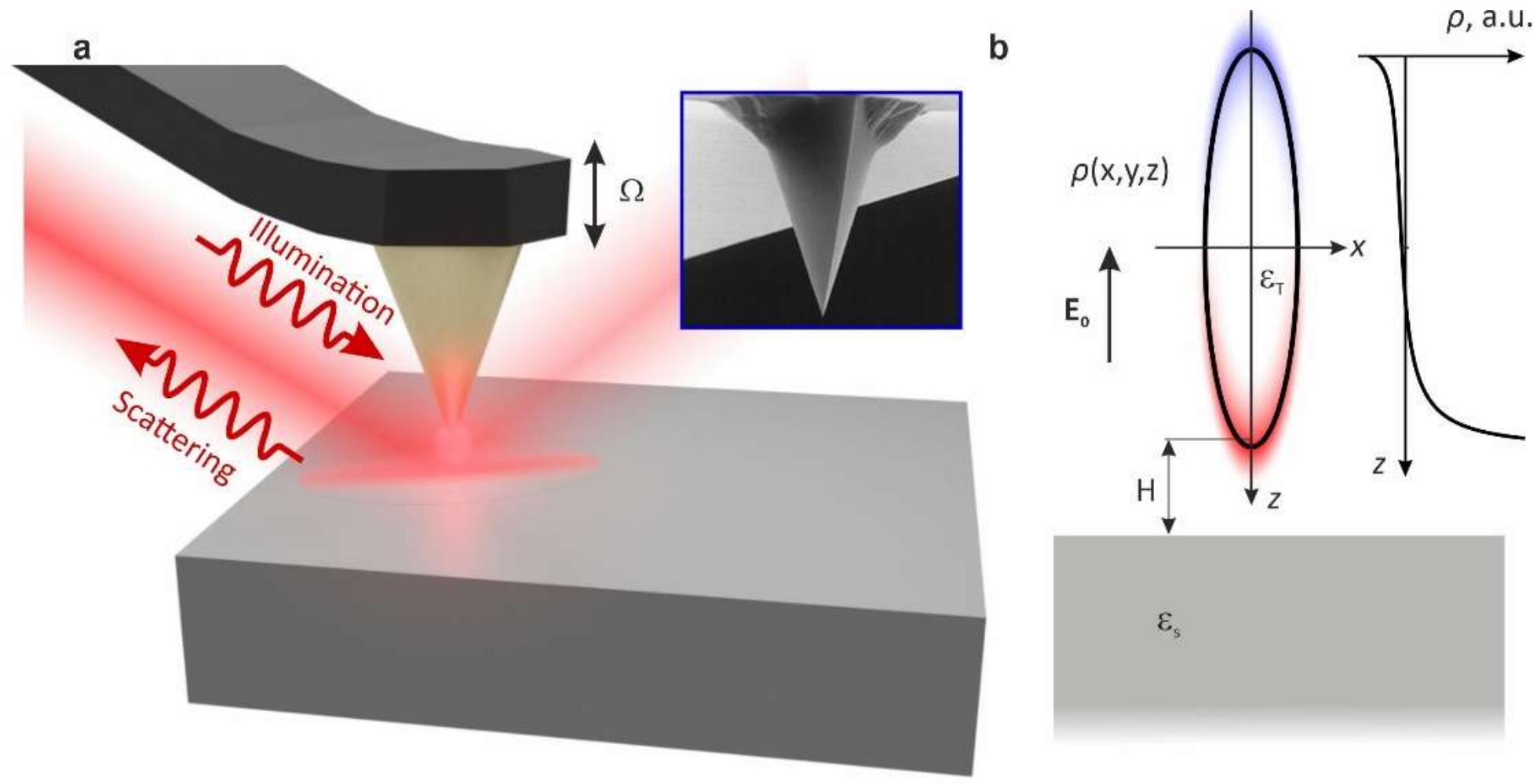

**Figure 1. a)** s-SNOM cantilever over the surface of a homogeneous bulk sample. Insert: SEM image of a typical s-SNOM tip made by NanoWorld®. **b)** Schematic of the charge distribution on the surface of the prolate spheroid placed in a uniform external field above the surface of the sample.

To validate the accuracy of our solution, we compare the results obtained using Eqs. (5)-(10) with full-wave numerical simulations based on the finite element method (FEM). As a test case, we consider a hypothetical material whose dielectric permittivity captures representative optical responses across a broad spectral range. Specifically, the material's permittivity includes a Drude term ($\omega_p = 500\ \mathrm{cm}^{-1}$, $\gamma_p = 150\ \mathrm{cm}^{-1}$, and $\varepsilon_\infty = 5$), a strong Lorentz term ($A_S = 1200\ \mathrm{cm}^{-1}$, $\omega_S = 1000\ cm^{-1}$, and $\gamma_S = 60\ \mathrm{cm}^{-1}$), and a weak Lorentz term ($A_W = 500\ \mathrm{cm}^{-1}$, $\omega_W = 1800\ \mathrm{cm}^{-1}$, and $\gamma_W = 30\ \mathrm{cm}^{-1}$), as illustrated in Fig. 2a. The spheroidal tip is modeled with a curvature radius $R = 25$ nm, a total length $L = 600$ nm, an oscillation amplitude $A = 50$ nm, and a minimum tip–sample distance $H_0 = 2$ nm. Figure 2b presents the amplitude and phase of the second and fourth harmonics of the polarizability $\alpha_{nz}$ of the spheroid placed above the hypothetical sample, normalized to that of the spheroid placed above the gold surface, $\alpha_{nzAu}$, calculated using Eq. (3). The first and third plot represent the amplitudes, $|\alpha_{nz}/\alpha_{nzAu}|$, and the second and fourth plots show the phases, $\arg(\alpha_{nz}/\alpha_{nzAu})$. The analytical and numerical results, blue line and blue dots respectively, exhibit excellent agreement across the entire frequency range, including regions with strong spectral features. This agreement, within the limits of numerical precision, confirms the validity and accuracy of the analytical solution. Furthermore, Fig. 2b includes results obtained using FDM, red dashed line, with a fitted complex coefficient $g$ [24], optimized to match the numerical results. A reported value of the parameter $g$ that can be used to obtain a qualitative result and as an initial approximation for quantitative spectrum fitting is $g = 0.7e^{0.06i}$ [24]. In our case, the best fit is achieved with $g = 0.75e^{0.029i}$. The FDM enables a computationally efficient estimation of the near-field response through the tuning of a single complex-valued parameter. However, the utility of the FDM for predictive or interpretive purposes is limited, as the parameter $g$ cannot be calculated or measured directly. In contrast, all parameters in our model are specific physical quantities corresponding to the parameters of a real tip, except the spheroid length, which we discuss later.

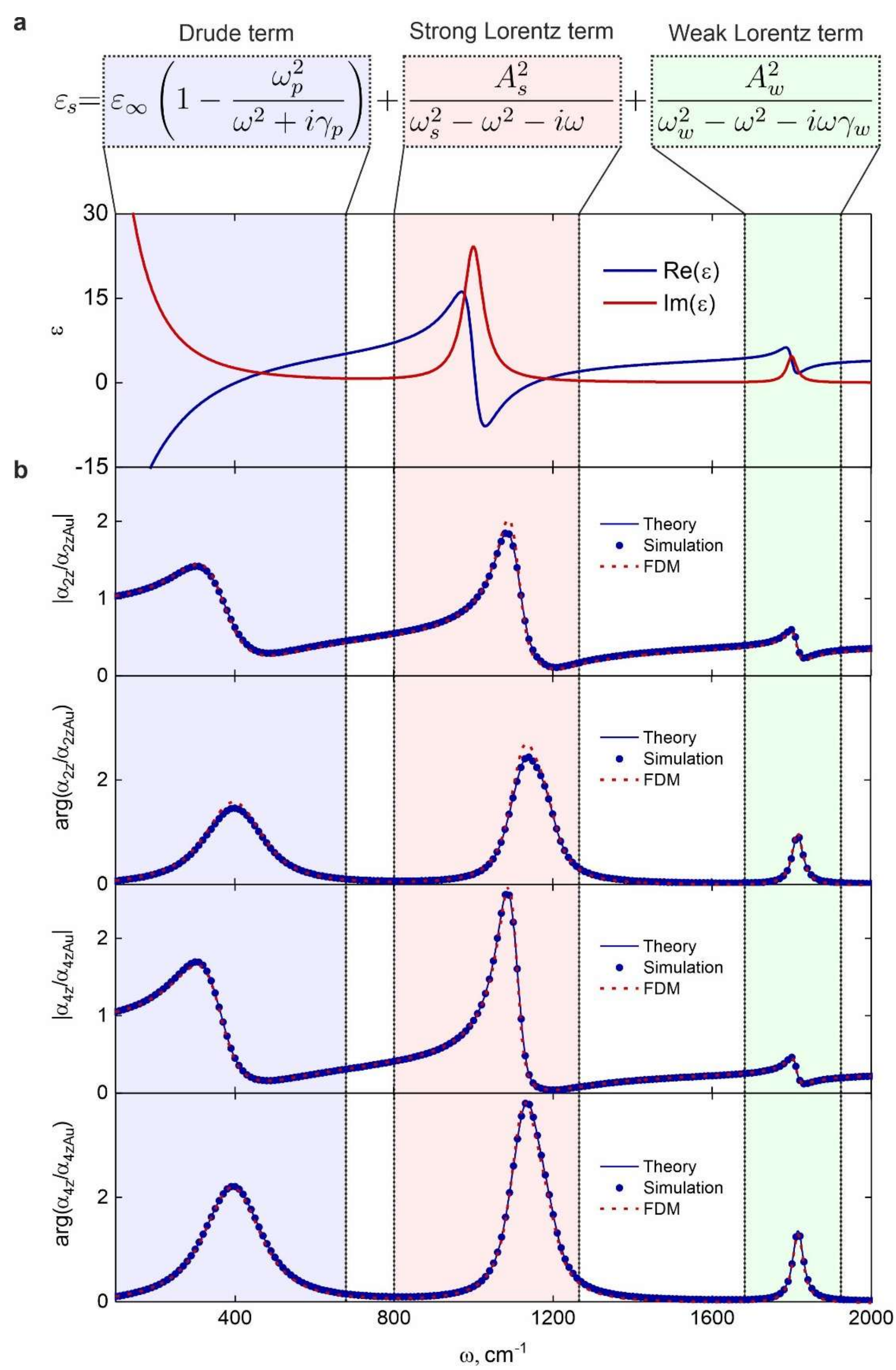

**Figure 2. a)** Dielectric permittivity function of a hypothetical sample material including Drude term ($\omega_p = 500\ \mathrm{cm}^{-1}$, $\gamma_p = 150\ \mathrm{cm}^{-1}$, and $\varepsilon_\infty = 5$), a strong Lorentz term ($A_S = 1200\ \mathrm{cm}^{-1}$, $\omega_S = 1000\ cm^{-1}$, and $\gamma_S = 60\ \mathrm{cm}^{-1}$), and a weak Lorentz term ($A_w = 500\ \mathrm{cm}^{-1}$, $\omega_w = 1800\ \mathrm{cm}^{-1}$, and $\gamma_w = 30\ \mathrm{cm}^{-1}$). **b)** The amplitude and phase of the second and fourth harmonics of the moving spheroid polarizability, calculated for the spheroid over the sample and normalized to those of the spheroid over the gold. The first and second panels from the top represent the normalized amplitude and phase of the second harmonic, $|\alpha_{2z}/\alpha_{2zAu}|$, and $\arg(\alpha_{2z}/\alpha_{2zAu})$, respectively. The third and fourth panels represent the normalized

amplitude and phase of the fourth harmonic. Solid curve represents analytically calculated data and dots corresponds to FEM simulation. The geometrical parameters of the model are $R = 25$ nm, $L = 600$ nm, $A = 50$ nm, $H_0 = 2$ nm. The simulated curve was fitted by FDM (dashed curve), assuming the same geometrical parameters; the obtained value of $g$-coefficient is $g = 0.75e^{0.029i}$.

The availability of an accurate quantitative model significantly expands the potential for predicting and analyzing s-SNOM near-field spectra. Such a model facilitates the interpretation of spectral features, enables the extraction of local optical properties via inverse modeling, and provides a robust framework for describing optical phenomena at the nanoscale. A key advantage of the model lies in its computational efficiency, which allows for rapid exploration of how various system parameters influence the measured signal. Among the model parameters unrelated to the intrinsic optical properties of the sample, the most critical are the spheroid's permittivity, curvature radius, total length, modulation amplitude, and the minimum distance between the spheroid apex and the sample surface. Of these, the minimal tip–sample distance, $H_0$, and the tip curvature radius, $R$, have the most pronounced impact on the near-field response. While the modulation amplitude also plays a significant role, it is typically relatively well-controlled and directly measurable in experiments. A comprehensive analysis of the sensitivity of the near-field spectrum to these parameters is provided in Supporting Information, Section 4. Importantly, although $H_0$ exerts a particularly strong influence on the near-field spectra, this parameter is challenging to determine experimentally, as it is governed by a complex interaction force between the tip and the sample. Improving the accuracy of this parameter's estimation would substantially enhance the predictive capability and reliability of the model, and is therefore an important experimental problem for future research.

Another parameter of interest is the total length of the prolate spheroid used in the model. This is the only parameter that does not correspond directly to any specific geometric parameter of the actual tip, which in reality exhibits a conical or pyramidal shape with a much greater extent. Nevertheless, while the spheroid length has an impact on the calculated spectra, its influence is comparatively minor relative to that of $R$ and $H_0$ (see Supporting Information, Section 4). As such, it can be interpreted as an effective model parameter that encapsulates contributions from the tip geometry, retardation effects, beam focusing parameters and other complex physical factors. Determining the appropriate value for this parameter remains an open problem for future investigation. At present, a detailed study of the spheroid length role in the performance of the model in comparison to experimental spectra is hindered by uncertainties in more influential parameters, most notably, $R$ and $H_0$, which dominate the near-field signal.

To demonstrate the applicability of the analytical spheroid model to experimental data, we considered near-field spectra of two representative materials: PMMA, which exhibits a weak Lorentzian feature in its dielectric function, and crystalline quartz, characterized by a strong Lorentz-type resonance. Here we measured the near-field spectra of quartz; the detailed information on the experimental setup and measurement procedures is provided in Supporting Information, Section 5. The PMMA spectrum is taken from the following source [52]. Figures 3a and 3b present the measured and simulated near-field spectra, $\sigma_n(\omega)$, calculated using Eq. (4), for PMMA and quartz, respectively. The PMMA spectra are normalized to the near-field signal from silicon, $\sigma_{nSi}(\omega)$, and the quartz spectra are normalized to the gold, $\sigma_{nAu}(\omega)$. To account for the finite spectral resolution of the experimental apparatus, the calculated quartz spectra are convolved with a Gaussian function having a full width at half maximum (FWHM) of 7 cm$^{-1}$, corresponding to the interferometer's bandwidth. In the analytical calculations, the amplitude of tip oscillation, $A$, is set equal to the same as in the corresponding experiment. A standard radius of curvature for s-SNOM tips of $R = 25$ nm [1, 25] is assumed. The spheroid length, $L$, serves as a fitting parameter. Additionally, since the minimum tip-sample distance, $H_0$, cannot be precisely measured experimentally, it is varied within a reasonable range of 0 to 3 nm to optimize the agreement between theoretical predictions and experimental spectra. The dielectric permittivity of the tip is set to the dielectric permittivity of platinum film [53]. The analytical model reproduces the PMMA spectrum with excellent fidelity. For quartz, the primary resonance is well captured, although a slight shift is observed in the high-frequency peak, likely attributable to discrepancies between the modeled and actual dielectric functions of the material. Overall, these results confirm the capability of the spheroid model to accurately describe near-field responses in realistic experimental conditions.

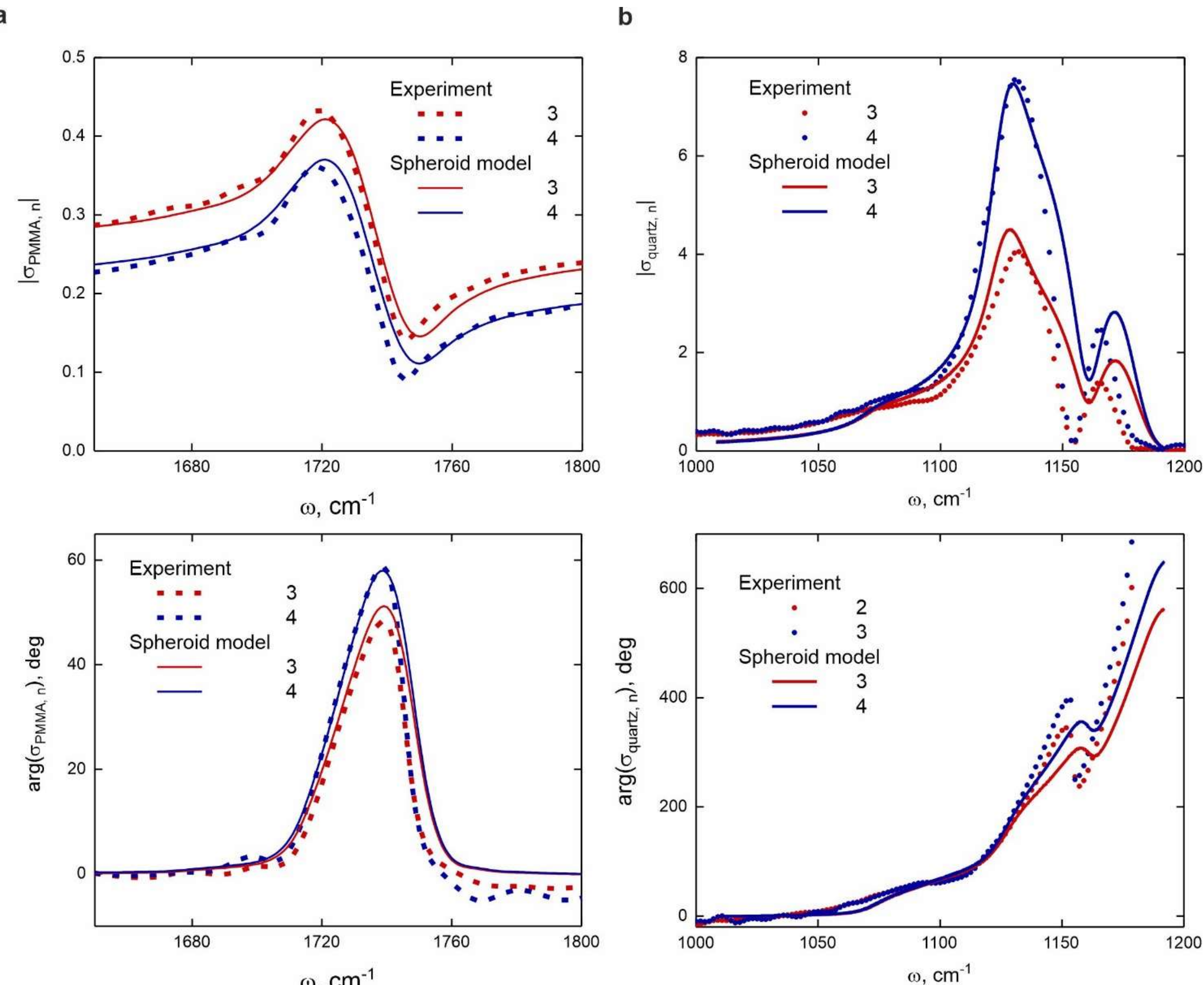

**Figure 3.** Comparison of the measured near-field spectra (the amplitude and phase of the normalized 3rd and 4th harmonics of the near-field signal, $\sigma_n$) with the spectra calculated by the spheroid model, dashed and solid curves, respectively. **a)** Spectra of PMMA normalized to Si, the geometrical parameters of the model are $R = 25$ nm, $A = 30$ nm, $L = 200$ nm, the minimal distance between the tip and the PMMA surface is $H_{0PMMA} = 2$ nm, the minimal distance between the tip and the silicon surface is $H_{0Si} = 2.8$ nm. **b)** Spectra of quartz normalized to gold, the geometrical parameters of the model are $R = 25$ nm, $A = 40$ nm, $L = 600$ nm, the minimal distance between the tip and the quartz surface is $H_{0SiO_2} = 2$ nm, the minimal distance between the tip and the gold surface is $H_{0Au} = 4.75$ nm.

## Conclusion

We have developed a quantitative analytical model for scattering-type scanning near-field optical microscopy (s-SNOM), based on a prolate spheroid approximation of the tip and the quasi-electrostatic limit. This model enables efficient and quantitatively reliable computation of near-field spectra for isotropic and uniaxial bulk materials, avoiding the need for empirical fitting parameters common in phenomenological models. Our model provides a robust framework for interpreting experimental data, as shown through comparisons with measured spectra of PMMA and quartz. Furthermore, it enables inverse reconstruction of the sample's dielectric function and facilitates systematic analysis of how geometric and material parameters influence the near-field signal. However, some geometric parameters, in particular the minimum distance between the tip and the sample, are usually not known with sufficient accuracy in current experimental setups and still remain in the role of fitting parameters of the model. More accurate methods for determining these parameters in the future will allow an even more significant increase and broaden the model's applicability. Owing to its speed and accuracy, the model also presents a valuable tool for generating synthetic datasets to train machine learning algorithms for advanced spectral analysis [54-56]. Nevertheless, the proposed model is developed and validated specifically for the mid-IR range and may not yield reliable quantitative results outside this region. In the visible range, the electrostatic approximation might fail due to the shorter wavelengths, while at terahertz frequencies, the longer wavelengths could induce antenna-like resonances in the tip, making tip geometry increasingly important. Thus, exploring the model's applicability and potential generalization beyond the mid-IR range remains an important topic for future research.

To promote further research and broader adoption, we have implemented the model in a publicly available numerical tool (see Supporting Information, Section 6, https://github.com/Voronin-Kirill/s-SNOM_spectra). We believe this approach will significantly advance both theoretical modeling and experimental interpretation in s-SNOM, contributing to deeper insights into nanoscale optical phenomena.

**Acknowledgements.** We thank Martin Schnell and Lars Mester for fruitful discussions and constructive suggestions for our work. We thank Edoardo Vicentini for his help with optimizing the calculations and translating the script from MATLAB to Python. K.V.V. received the support of a fellowship from “la Caixa” Foundation (ID 100010434) with the fellowship code LCF/BQ/DI21/11860026. A.Y.N acknowledges support from the Spanish Ministry of Science and Innovation (grant PID2023-147676NB-I00) and the Basque Department of Education (grant PIBA-2023-1-0007).

# Supporting Information for "Quantitative Analytical Model for Scattering-type Scanning Near-field Optical Spectroscopy"

Kirill V. Voronin[1,2], Iker Herrero León[3], Rainer Hillenbrand[3,4], Alexey Y. Nikitin[1,4,†]

[1]Donostia International Physics Center (DIPC), Donostia-San Sebastián 20018, Spain
[2]Universidad del Pais Vasco/Euskal Herriko Unibertsitatea, 20080 Donostia/San Sebastián, Basque Country, Spain
[3]CIC nanoGUNE BRTA, Tolosa Hiribidea 76, 20018 Donostia-San Sebastian, Spain
[4]IKERBASQUE, Basque Foundation for Science, Bilbao 48013, Spain

[†]Contact author: alexey@dipc.org

## S1. PROBLEM STATEMENT. SCATTERED FIELD AND DIPOLE MOMENT

The problem of determining the electromagnetic field scattered by the s-SNOM tip can be effectively separated into two parts: the far-field and near-field problems. The far-field analysis involves computing the incident wave field near the apex of the tip and determining the scattered wave field resulting from the induced charge distribution on the tip's surface. In contrast, the near-field analysis centers on calculating the surface charge density on the tip when positioned above the sample surface under an external electromagnetic excitation (Fig. S1).

The principal complexity arises predominantly in the near-field analysis. Here, we employ an electrostatic approximation suitable for conditions where the characteristic length scales—specifically, the tip radius $R$ and the tip-to-surface separation distance $H$—are significantly smaller than the incident radiation wavelength $\lambda$. Within this electrostatic approximation, we neglect the vector potential of the electromagnetic field as well as retardation effects at these length scales, effectively simplifying the problem to an electrostatic one of determining the electric potential distribution. Consequently, the incident wave field in this near-field region can be considered as a spatially uniform background electric field, denoted by $\mathbf{E}_b$. Furthermore, given the geometry of an elongated, sharp tip, we simplify the analysis by considering only the vertical component of this incident electric field, designated for brevity as $E_0 \equiv E_{b,z}$, since this component predominantly influences the surface charge redistribution.

Focusing now on the far-field part of the analysis, we observe that the uniform background electric field, $\mathbf{E}_b$, resulting from the interference between the incident and reflected plane waves at the substrate surface (Fig. S1), can be expressed as:

$$E_{bz} = E_{inc}\cos\theta + E_{ref}\cos\theta = E_{inc}(1 + r_p)\cos\theta, \tag{S1}$$

where $\theta$ is the angle of incidence, $E_{inc}$ and $E_{ref}$ denote the amplitudes of the incident and reflected electric fields, respectively, and $r_p$ represents the Fresnel reflection coefficient for p-polarized radiation.

To determine the field scattered by the tip, we utilize the approximation that, due to demodulation procedures employed in s-SNOM measurements, only the contributions from charges localized at subwavelength scales near the tip apex are relevant. At far-field distances, these charges effectively behave as a dipole source characterized by the dipole moment $p_z$, resulting in a scattered electric field of the form:

$$E_{sca} \sim (1 + r_p)p_z, \tag{S2}$$

where the factor $(1 + r_p)$ accounts for both the direct and substrate-reflected contributions to the scattered wave generated by the induced dipole. To extract the signal component arising specifically from the near-field interaction between the tip and the sample, the tip-sample distance, $H$, is modulated at frequency $\Omega$. This modulation enables the isolation of the distance-dependent signal component. Subsequently, the scattered wave is combined with a reference beam, $E_{ref}$, modulated in phase at frequency $M$, chosen not to be a multiple of $\Omega$, and detected. The resulting electrical signal at the detector, proportional to the incident radiation intensity, can be expressed as follows

$$U \propto I_d = \left|E_{sca} + E_{ref}\right|^2 = |E_{sca}|^2 + 2\mathrm{Re}\left(E_{sca}E_{ref}^*\right) + \left|E_{ref}\right|^2 \tag{S3}$$

The term proportional to $E_{sca}$ ($2\mathrm{Re}\left(E_{sca}E_{ref}^*\right)$) consists of harmonics with frequencies $n\Omega + mM$. By isolating these harmonics, both amplitude and phase information of $E_{sca}$ can be obtained [1].

Introducing the tip polarizability $\alpha_z = p_z/E_0$, the normalized scattered field amplitude becomes:

$$\frac{E_{sca}}{E_{inc}} \sim (1 + r_p)^2\alpha_z. \tag{S4}$$

Thus, the scattering problem ultimately reduces to computing the tip's induced dipole moment. Finally, to more accurately represent realistic experimental conditions, the prefactor $(1 + r_p)^2$ can be further adjusted to incorporate sample-specific heterogeneities such as proximity to edges, local focusing effects, or other perturbations influencing the interaction between direct and reflected fields as follows

$$\frac{E_{sca}}{E_{inc}} \sim (1 + \varsigma r_p)^2\alpha_z. \tag{S5}$$

where $\varsigma$ is some unknown weight coefficient [2].

To determine the dipole moment induced on the tip positioned in a uniform electric field, we solve Laplace's equation for the scalar potential under appropriate boundary conditions. These conditions include continuity of the electric potential and continuity of the normal component of the electric displacement field, **D**, across both the tip and sample surfaces. Employing the superposition principle allows simplification of the problem by decomposition of the field into distinct contributions as follows:

$$\mathbf{E}(\mathbf{r}) = \mathbf{E}_b + \mathbf{E}_t(\mathbf{r}) + \mathbf{E}_s(\mathbf{r}), \tag{S6}$$

where $\mathbf{E}(\mathbf{r})$ represents the total electric field distribution around the tip, $\mathbf{E}_b$ is the background electric field formed by the incident and reflected plane waves (assumed uniform), $\mathbf{E}_t(\mathbf{r})$ is the field generated by charges induced on the tip, and $\mathbf{E}_s(\mathbf{r})$ is the field arising from charges induced within the substrate due to the presence of the tip. Notably, $\mathbf{E}_b$, which incorporates both incident and reflected plane waves (Fig. S1), inherently satisfies boundary conditions at the substrate surface.

To further simplify the analysis, we introduce the surface charge density on the tip, $\rho(\mathbf{r})$, and separate it into two distinct components, $\rho(\mathbf{r}) = \rho_0(\mathbf{r}) + \rho_i(\mathbf{r})$. The first component, $\rho_0(\mathbf{r})$, defines a field $\mathbf{E}_{t0}(\mathbf{r})$ which, when combined with the background field $\mathbf{E}_b$, satisfies boundary conditions at the tip surface. The second component, $\rho_i(\mathbf{r})$, defines the field $\mathbf{E}_{ti}(\mathbf{r})$ which, combined with the field $\mathbf{E}_s(\mathbf{r})$, also meets the tip surface boundary conditions. The field $\mathbf{E}_s(\mathbf{r})$ is calculated via the electrostatic image method applicable to an infinite planar substrate to fulfill boundary conditions at the sample surface. Consequently, Eq. (S6) can be reformulated as:

$$\mathbf{E}(\mathbf{r}) = \left(\mathbf{E}_b + \mathbf{E}_{t0}(\mathbf{r})\right) + \left(\mathbf{E}_{ti}(\mathbf{r}) + \mathbf{E}_s(\mathbf{r})\right). \tag{S7}$$

The field described by Eq. (S7) satisfies Maxwell's equations and all relevant boundary conditions. Specifically, the fields in each bracket individually satisfy boundary conditions at the tip surface, while the combined field $\mathbf{E}_{t0}(\mathbf{r}) + \mathbf{E}_{ti}(\mathbf{r}) + \mathbf{E}_s(\mathbf{r})$ fulfills boundary conditions at the substrate surface.

Thus, the charge distribution $\rho_0(\mathbf{r})$ depends solely on the background electric field and tip geometry, whereas distribution $\rho_i(\mathbf{r})$ depend on $\rho_0(\mathbf{r})$, tip geometry, and the tip-to-substrate distance, $H$. Consequently, the charge distribution $\rho_0(\mathbf{r})$ does not directly contribute to the non-zero harmonic components of the detected signal; however, its computation is an essential intermediate step in the analysis.

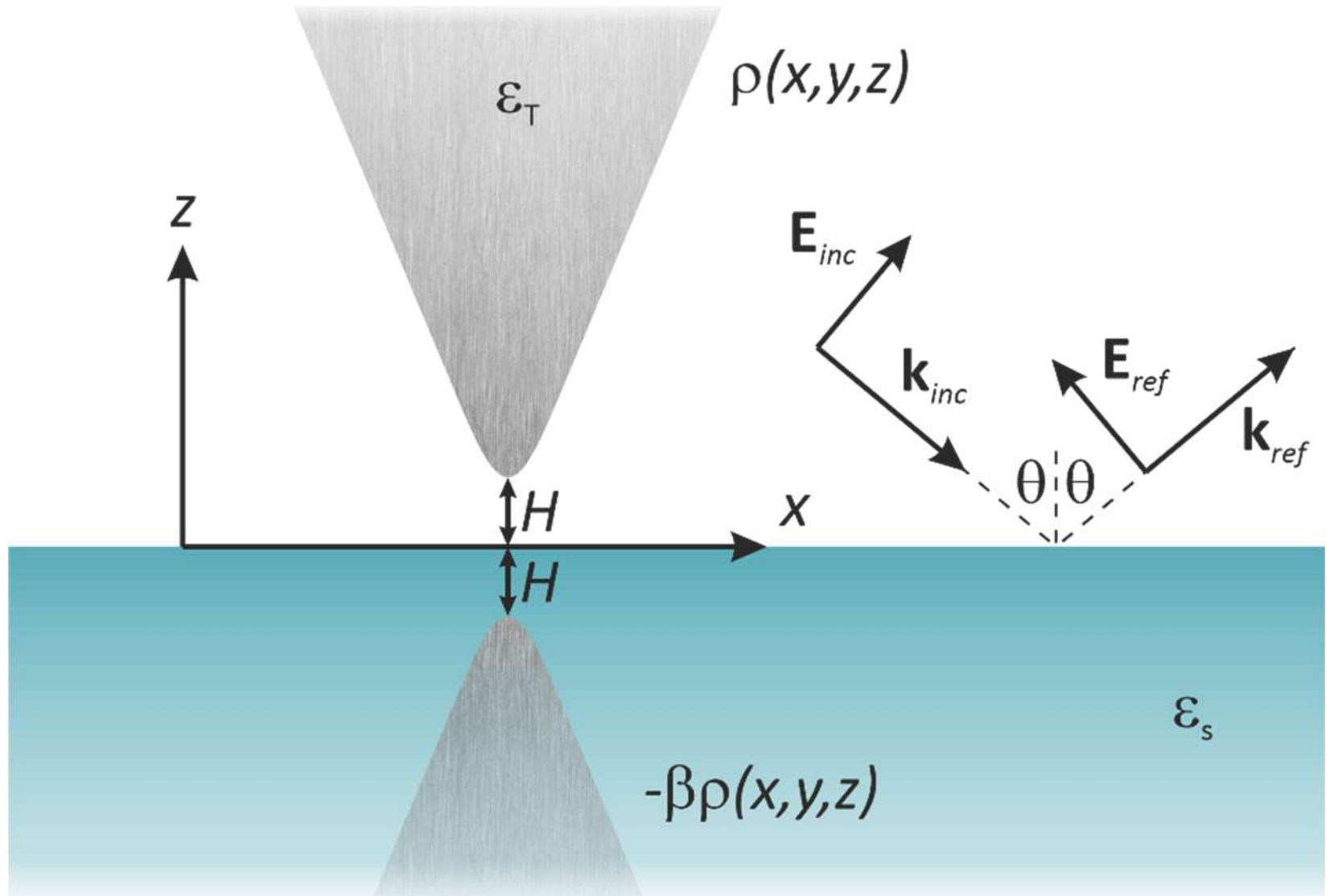

FIG. S1. Schematic of the tip with the dielectric permittivity $\varepsilon_T$ placed over the sample with the dielectric permittivity $\varepsilon_s$ irradiated by the plane wave with the amplitude $\mathbf{E}_{inc}$ at the angle $\theta$. $H$ is the tip-sample distance; $\rho(x, y, z)$ is the charge density distribution at the tip surface. The field of the sample can be modeled by the field of the image tip with the surface charge density $-\beta\rho(x,y,z)$.

## S2. ELECTROSTATIC IMAGE METHOD FOR SAMPLE SURFACE

The electrostatic image method for an infinite planar interface can be derived from the Green's function for Poisson's equation within a uniform medium characterized by dielectric permittivity $\varepsilon_1$:

$$G(\mathbf{r}-\mathbf{r}') = \iiint_{\mathbb{R}^3} \frac{d^3\mathbf{k}}{(2\pi)^3}\frac{4\pi}{\varepsilon_1 k^2} e^{i\mathbf{k}(\mathbf{r}-\mathbf{r}')} = \iint_{\mathbb{R}^2} \frac{d^2\mathbf{k}_\parallel}{2\pi\varepsilon_1 k_\parallel} e^{i\mathbf{k}_\parallel(\mathbf{r}_\parallel-\mathbf{r}'_\parallel)-k_\parallel|z-z'|}. \quad \text{(S8)}$$

Considering a point charge $q$ located at $(0,0,z_0)$ above the plane defined by $z=0$, the electric potential generated by this charge is expressed as:

$$\varphi(\mathbf{r}) = \frac{q}{\varepsilon_1}\left(\iint_{\mathbb{R}^2} \frac{d^2\mathbf{k}_\parallel}{2\pi k_\parallel} e^{i\mathbf{k}_\parallel(\mathbf{r}_\parallel-\mathbf{r}'_\parallel)-k_\parallel|z-z_0|} + \iint_{\mathbb{R}^2} \frac{d^2\mathbf{k}_\parallel}{2\pi k_\parallel} r_e(\mathbf{k}_\parallel) e^{i\mathbf{k}_\parallel(\mathbf{r}_\parallel-\mathbf{r}'_\parallel)-k_\parallel(z+z_0)}\right). \quad \text{(S9)}$$

where $r_e(\mathbf{k})$ represents the electrostatic reflection coefficient associated with electric potential harmonics. By applying the boundary conditions for continuity of both electric field **E** and displacement field **D**, it can be demonstrated that this electrostatic reflection coefficient corresponds to the reflection coefficient of p-polarized waves in the high-wavevector limit: $r_e(\mathbf{k}_\parallel) = r_p(\mathbf{k}_\parallel)|_{k_0\varepsilon \ll k_\parallel}$.

For instance, considering a semi-infinite substrate with dielectric permittivity $\varepsilon_2$, the reflection coefficient simplifies in the large wavevector limit to:

$$r_e(\mathbf{k}_\parallel) \underset{\mathbf{k}_\parallel\to\infty}{\to} \frac{\varepsilon_1-\varepsilon_2}{\varepsilon_1+\varepsilon_2} = -\beta_i \quad \text{(S10)}$$

Consequently, the potential becomes:

$$\varphi(\mathbf{r}) = \frac{q}{\varepsilon_1}\left(\iint_{\mathbb{R}^2} \frac{d^2\mathbf{k}_\parallel}{2\pi k_\parallel} e^{i\mathbf{k}_\parallel(\mathbf{r}_\parallel-\mathbf{r}'_\parallel)-k_\parallel|z-z_0|} - \beta_i \iint_{\mathbb{R}^2} \frac{d^2\mathbf{k}_\parallel}{2\pi k_\parallel} e^{i\mathbf{k}_\parallel(\mathbf{r}_\parallel-\mathbf{r}'_\parallel)-k_\parallel(z+z_0)}\right), \quad \text{(S11)}$$

The first integral in Eq. (S11) corresponds to the potential of a charge $q$ in free space located at $(0,0,z_0)$, while the second integral represents the potential due to an image charge $-\beta_i q$ placed at $(0,0,-z_0)$, thereby recovering the standard electrostatic image solution for a semi-infinite dielectric medium.

Analogously, for a uniaxial medium with optical axis perpendicular to the interface, the reflection coefficient is:

$$r_e(\mathbf{k}_\parallel) \underset{\mathbf{k}_\parallel\to\infty}{\to} \frac{\varepsilon_1-\sqrt{\varepsilon_\perp\varepsilon_\parallel}}{\varepsilon_1+\sqrt{\varepsilon_\perp\varepsilon_\parallel}} = -\beta_u \quad \text{(S12)}$$

Hence, for a uniaxial dielectric half-space, the image charge corresponding to a point charge $q$ located at $(0,0,z_0)$ is given by $-\beta_u q$ and located at $(0,0,-z_0)$.

## S3. DERIVATION OF THE ANALYTICAL EXPRESSION FOR THE DIPOLE MOMENT OF THE SPHEROID

In this section, we calculate the surface charge distribution of a prolate spheroid with the dielectric permittivity $\varepsilon_T$ positioned above a planar surface of a homogeneous semi-infinite medium under the influence of a uniform external electric field (Fig. S2), aiming ultimately to determine the resulting dipole moment.

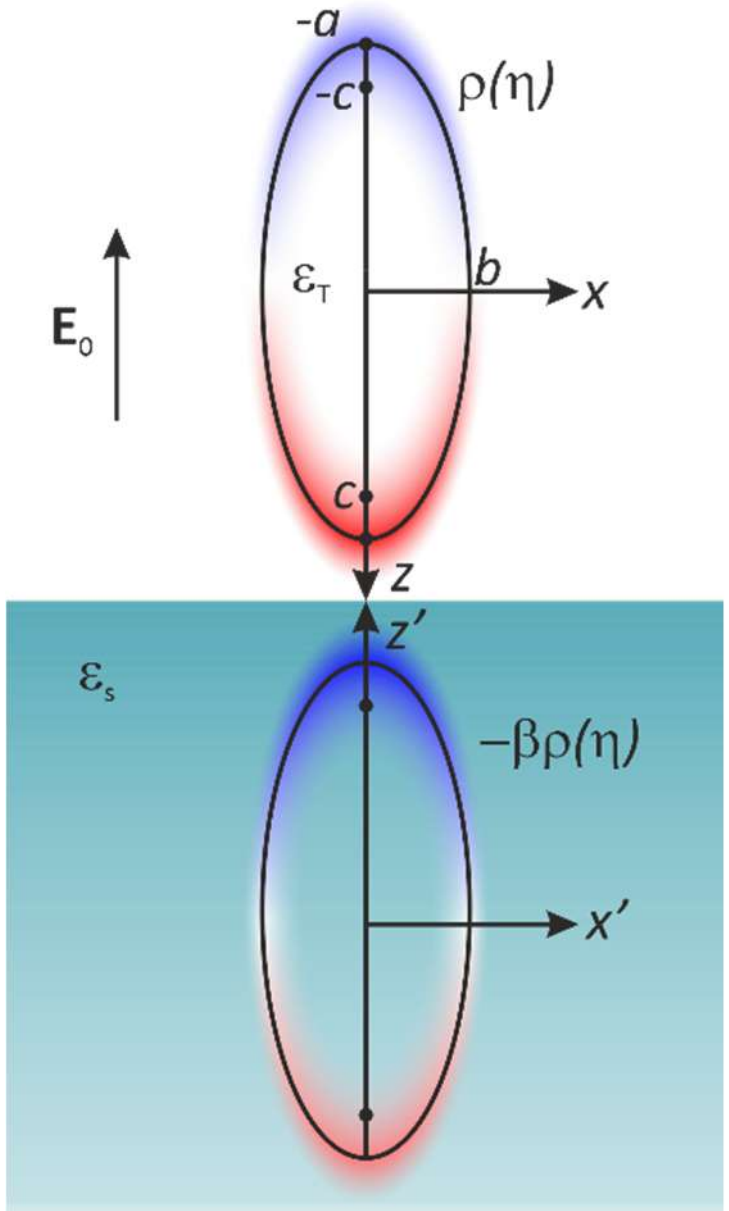

FIG. S2. Schematic of the spheroid with the dielectric permittivity $\varepsilon_T$ placed over the sample with the dielectric permittivity $\varepsilon_s$ in the uniform vertical electric field $\mathbf{E}_0$. $a$ and $b$ are major and minor spheroid semi-axes respectively; $c$ and $-c$ are the spheroid foci positions. $\rho(\eta)$ is the charge density distribution at the spheroid surface. The field of the sample can be modeled by the field of the image spheroid with the surface charge density $-\beta\rho(\eta)$.

To facilitate the analysis, we reposition the origin of the coordinate system to coincide with the center of the spheroid (Fig. S2). We then introduce the prolate spheroidal coordinates $(\mu, \nu, \phi)$, defined as follows:

$$\begin{aligned} x &= c\sinh\mu\sin\nu\cos\phi, \\ y &= c\sinh\mu\sin\nu\sin\phi, \\ z &= c\cosh\mu\cos\nu, \end{aligned} \tag{S13}$$

where $c$ represents the distance from the spheroid's center to its focal point, and the coordinate variables span $\mu \in [0, \infty)$, $\nu \in [0, \pi]$, and $\phi \in [0, 2\pi]$. Within this coordinate framework, the spheroid surface corresponds to $\mu = \mu_0$. The major semi-axis length $a = L/2$ is related to $\mu_0$ by $a = c\cosh\mu_0$, while the minor semi-axis length is given by $b = c\sinh\mu_0$. The radius of curvature at the tip apex can be directly determined from these geometrical relations as follows:

$$R = b^2/a = c\sinh\mu_0\tanh\mu_0 \tag{S14}$$

### A. General expression for the electric potential inside and outside of the spheroid and the charge distribution at the spheroid surface

Laplace's equation expressed in prolate spheroidal coordinates takes the following form [3]:

$$\Delta\varphi(\xi, \eta) = \frac{1}{\xi^2 - \eta^2}\left[\frac{\partial}{\partial\xi}(\xi^2 - 1)\frac{\partial}{\partial\xi} + \frac{\partial}{\partial\eta}(1 - \eta^2)\frac{\partial}{\partial\eta} + \frac{\xi^2 - \eta^2}{(\xi^2 - 1)(1 - \eta^2)}\frac{\partial^2}{\partial\phi^2}\right]\varphi_r(\xi, \eta) = 0. \tag{S15}$$

where we introduced the notation $\xi = \cosh\mu$ and $\eta = \cos\nu$. The general solution to this equation is provided by [4]:

$$\varphi_\kappa^\zeta(\xi, \eta) = \left[A_\kappa^\zeta P_\kappa^\zeta(\xi) + B_\kappa^\zeta Q_\kappa^\zeta(\xi)\right]\left[C_\kappa^\zeta P_\kappa^\zeta(\eta) + D_\kappa^\zeta Q_\kappa^\zeta(\eta)\right]\left[E_\kappa^\zeta \sin\zeta\phi + F_\kappa^\zeta \cos\zeta\phi\right]. \tag{S16}$$

where $P_\kappa^\zeta(x)$ and $Q_\kappa^\zeta(x)$ represent the Legendre functions of the first and second kind, respectively, with degree $\kappa$ and order $\zeta$.

In this study, we focus on axially symmetric systems with respect to the $z$-axis; thus, the potential $\varphi$ is independent of the angular coordinate $\phi$ (which is equivalent to $\zeta = 0$), reducing Eq. (S16) to the simplified form:

$$\varphi_\kappa(\xi,\eta) = [A_\kappa P_\kappa(\xi) + B_\kappa Q_\kappa(\xi)][C_\kappa P_\kappa(\eta) + D_\kappa Q_\kappa(\eta)]. \tag{S17}$$

To determine the constants in Eq. (S17), we recall that $Q_\kappa(x) \to \infty$ as $x \to 1$ for all $\kappa$, whereas $P_\kappa(x) \to \infty$ as $x \to -1$ for non-integer $\kappa$ [5]. Therefore, $D_\kappa = 0$ and $\kappa = n = 0,1,2,..,$ as $\eta$ varies between $-1$ and 1 both inside and outside the spheroid. Furthermore, for large arguments, $P_n(x) \to \infty$ and $Q_n(x) \to 0$. Thus, the potential outside the spheroid, which remains finite at infinity, is given by:

$$\varphi_n(\xi,\eta) = C_n Q_n(\xi) P_n(\eta). \tag{S18}$$

while the potential inside the spheroid is:

$$\varphi_n(\xi,\eta) = C_n P_n(\xi) P_n(\eta). \tag{S19}$$

The potential outside the spheroid, $\varphi_{\text{out}}$, is represented as a sum of the reflected potential induced by the spheroid ($\varphi_r$) and the external potential ($\varphi_e$), accounting for the background field and induced charges on the sample surface. The potentials satisfy the following boundary conditions at the spheroid surface ($\xi = \xi_0$):

$$\varphi_{\text{in}}(\xi_0,\eta) = \varphi_{\text{out}}(\xi_0,\eta) = \varphi_e(\xi_0,\eta) + \varphi_r(\xi_0,\eta) \tag{S20}$$

$$\varepsilon \frac{\partial}{\partial \xi} \varphi_{\text{in}}(\xi,\eta)\bigg|_{\xi_0} = \frac{\partial}{\partial \xi} \varphi_{\text{out}}(\xi,\eta)\bigg|_{\xi_0} \tag{S21}$$

Here Eq. (S20) corresponds to the continuity of the electric potential, and Eq. (S21) corresponds to the continuity of the normal component of the field $\mathbf{D} = -\varepsilon \nabla \varphi$.

To determine the surface charge distribution corresponding to the reflected potential $\varphi_r(\xi,\eta,\phi)$, we use the known potential of a point charge $q$ positioned at $(\xi_q,\eta_q,\phi_q)$ [3]:

$$\varphi_q(\xi,\eta,\phi) = \frac{q}{c} \sum_{n=0}^{\infty} (2n+1) \sum_{m=0}^{n} \epsilon_m\, i^m \left[\frac{(n-m)!}{(n+m)!}\right] \cos m(\phi - \phi_q) P_n^m(\eta_q) P_n^m(\eta) \begin{cases} P_n^m(\xi_q) Q_n^m(\xi); & \xi > \xi_q \\ P_n^m(\xi) Q_n^m(\xi_q); & \xi < \xi_q \end{cases}, \tag{S22}$$

where $\epsilon_m = 2 - \delta_{0m}$.

As we consider the problem with cylindrical symmetry, the charge density at the spheroid surface, $\rho(\eta,\phi)$, does not depend on $\phi$, that is, $\rho(\eta,\phi) = \rho(\eta)$. According to Eq. (S22), the potential created by the charged surface of prolate spheroid outside of the spheroid is given by the following expression:

$$\varphi_r(\xi,\eta,\phi) = \int_{-1}^{1} h_\eta\, d\eta_0 \int_0^{2\pi} h_\phi\, d\phi_0 \frac{\rho(\eta_0)}{c} \sum_{n=0}^{\infty} (2n+1) \sum_{m=0}^{n} \epsilon_m\, i^m \left[\frac{(n-m)!}{(n+m)!}\right] \times \\ \times \cos m(\phi - \phi_0) P_n^m(\eta_0) P_n^m(\eta) P_n^m(\xi_0) Q_n^m(\xi), \tag{S23}$$

where the expressions for Lame coefficients read:

$$h_\eta = c\sqrt{\frac{\xi^2 - \eta^2}{1 - \eta^2}}, \qquad h_\phi = c\sqrt{(\xi^2 - 1)(1 - \eta^2)}. \tag{S24}$$

Integrating over $\phi_0$ we obtain

$$\varphi_r(\xi,\eta,\phi) = 2\pi c \sum_{n=0}^{\infty}(2n+1)\, P_n(\eta)P_n(\xi_0)Q_n(\xi) \int_{-1}^{1}\sqrt{(\xi_0^2-\eta_0^2)(1-\eta_0^2)}\, P_n(\eta_0)\rho(\eta_0)d\eta_0. \tag{S25}$$

Notice that the Legendre polynomials are defined as an orthogonal system over the interval $[-1,1]$, orthogonality condition reads:

$$\int_{-1}^{1} P_n(x)P_m(x)dx = \frac{2\delta_{mn}}{2n+1}, \tag{S26}$$

where $\delta_{mn}$ is the Kronecker delta. Exploiting the orthogonality of Legendre polynomials, the charge density can be expanded as:

$$\rho(\eta) = \frac{\sum_{n=0}^{\infty}(2n+1)\upsilon_n P_n(\eta)}{2\pi c\sqrt{(\xi_0^2-1)(\xi_0^2-\eta^2)}}. \tag{S27}$$

The choice of the expansion into a series of Legendre polynomials is motivated by the form of the potential $\varphi_q(\xi,\eta)$ created outside of the spheroid by the point charge, $q$ placed at the spheroid surface, given by Eq. (S22). From the Eq. (S25) we derive the following expression for $\varphi_r(\xi,\eta)$:

$$\varphi_r(\xi,\eta) = 2\sum_{n=0}^{\infty}(2n+1)\,\upsilon_n P_n(\xi_0)P_n(\eta)Q_n(\xi). \tag{S28}$$

### B. Spheroid in the uniform external field

We begin by considering a vertically polarized uniform electric field, defined as $\mathbf{E}_b = (0,0,-E_0)$, as illustrated in Fig. S2. The corresponding background electric potential, denoted by $\varphi_e = \varphi_0$, is expressed as:

$$\varphi_0 = zE_0 = cE_0\cosh\mu\cos\nu = cE_0\xi\eta = cE_0P_1(\xi)P_1(\eta) \tag{S29}$$

Inside the spheroid, the potential can be represented using Eq. (S16) as follows:

$$\varphi_{\text{in}}(\xi,\eta) = 2\sum_{n=0}^{\infty}(2n+1)\,U_n P_n(\eta)P_n(\xi), \tag{S30}$$

Similarly, the reflected potential, as described by Eq. (S28), takes the form:

$$\varphi_r^{(0)}(\xi,\eta) = 2\sum_{n=0}^{\infty}(2n+1)\,\upsilon_n^{(0)} P_n(\xi_0)P_n(\eta)Q_n(\xi). \tag{S31}$$

Applying the boundary condition at the spheroid surface, $\varphi_{\text{in}}(\xi_0) = \varphi_{\text{out}}(\xi_0)$, we obtain:

$$2\sum_{n=0}^{\infty}(2n+1)\,\upsilon_n^{(0)} P_n(\xi_0)Q_n(\xi_0)P_n(\eta) + cE_0P_1(\xi_0)P_1(\eta) = 2\sum_{n=0}^{\infty}(2n+1)\,U_n^{(0)} P_n(\eta)P_n(\xi_0). \tag{S32}$$

From Eq. (S32) we obtain that $\upsilon_n^{(0)} = 0$ and $U_n^{(0)}$ for $n \neq 1$, therefore, we obtain

$$6\upsilon_1^{(0)}Q_1(\xi_0) + cE_0 = 6U_1^{(0)}. \tag{S33}$$

On the other hand, the second boundary condition, Eq. (S21), provides an additional equation for $\upsilon_1^{(0)}$ and $U_1^{(0)}$:

$$6\upsilon_1^{(0)}\xi_0\frac{\xi_0Q_1(\xi_0)-Q_0(\xi_0)}{\xi_0^2-1} + cE_0 = 6\varepsilon U_1^{(0)}. \tag{S34}$$

where we used the derivatives of the Legendre functions:

$$\frac{d}{dx}P_n(x) = \frac{n\xi_0P_n(\xi_0)-nP_{n-1}(\xi_0)}{\xi_0^2-1} \tag{S35}$$

$$\frac{d}{dx}Q_n(x) = \frac{n\xi_0 Q_n(\xi_0) - nQ_{n-1}(\xi_0)}{\xi_0^2 - 1}. \tag{S36}$$

From Eqs. (S33) and (S34), we finally derive:

$$\upsilon_1^{(0)} = -\frac{(\varepsilon-1)cE_0}{6\left(\varepsilon Q_1(\xi_0) - \xi_0 Q_0(\xi_0) + \frac{\xi_0^2}{\xi_0^2-1}\right)}. \tag{S37}$$

Consequently, the surface charge density on the spheroid is expressed by:

$$\rho_0(\eta) = \frac{3\upsilon_1^{(0)} P_1(\eta)}{2\pi c\sqrt{(\xi_0^2-1)(\xi_0^2-\eta^2)}} = -\frac{(\varepsilon-1)E_0\eta}{4\pi\left(\varepsilon Q_1(\xi_0) - \xi_0 Q_0(\xi_0) + \frac{\xi_0^2}{\xi_0^2-1}\right)\sqrt{(\xi_0^2-1)(\xi_0^2-\eta^2)}}. \tag{S38}$$

Equation Eq. (S38) brings us to the expression for the reflected potential for the perfectly conducting spheroid [6]:

$$\varphi_r^{(0)}(\xi,\eta) = -cE_0 P_1\left(\frac{a}{c}\right) P_1(\eta)\frac{Q_1(\xi)}{Q_1\left(\frac{a}{c}\right)} = -aE_0\eta\frac{2+\xi\log\frac{\xi-1}{\xi+1}}{2+\frac{a}{c}\log\frac{a-c}{a+c}}. \tag{S39}$$

## C. Spheroid in the field of image charges

Finally, we address the contribution to the potential arising from the sample charges, corresponding to the field $\mathbf{E}_s = -\nabla\varphi_s$ (Eq. (S6)). According to the electrostatic image method detailed in Section 2, within the region above the sample surface, this potential is equivalent to that generated by the image of the spheroid located symmetrically to the real spheroid relative to the surface and with a charge distribution equal to $-\beta\left(\rho_0(\eta') + \rho_i(\eta')\right)$, where $\eta'$ denotes the coordinate in the coordinate system of the spheroid image, $\rho_0$ represents the charge distribution induced by the external field potential $\varphi_0$, and $\rho_i$ denotes the charge distribution resulting from the image charge potential, $\varphi_s$.

Employing Eq. (S22), this potential can be calculated explicitly as:

$$\begin{aligned}\varphi_s(\xi,\eta,\phi) = -\beta\int_{-1}^{1} c^2\sqrt{(\xi_0^2-1)(\xi_0^2-\eta'^2)}\,d\eta'\int_0^{2\pi} d\phi'\,\frac{\rho_i(\eta')+\rho_0(\eta')}{c}\sum_{n=0}^{\infty}(2n+1)\times \\ \times\sum_{m=0}^{n}\epsilon_m\, i^m\left[\frac{(n-m)!}{(n+m)!}\right]\cos m(\phi+\phi')P_n^m(\eta)P_n^m(\eta_q)P_n^m(\xi)Q_n^m(\xi_q),\end{aligned} \tag{S40}$$

where $\eta_q$ and $\xi_q$ are the coordinates of the spheroid image in the coordinate system of the real spheroid. The relationship between $\left(\xi_q,\eta_q\right)$ and $(\xi',\eta')$ (where $\xi' = \xi_0$) is established in Cartesian coordinates as illustrated in Fig. S2:

$$c\xi_q\eta_q = z = 2a + 2H - z' = 2a + 2H - c\xi_0\eta', \tag{S41}$$

$$c\sqrt{\xi_q^2-1}\sqrt{1-\eta_q^2} = x = c\sqrt{\xi_0^2-1}\sqrt{1-\eta_q'^2}. \tag{S42}$$

Consequently, we derive the following relations:

$$\eta_q = \frac{\psi}{\xi_q}, \qquad \xi_q = \sqrt{\frac{1+\psi^2+\chi^2+\sqrt{(1+\psi^2+\chi^2)^2-4\psi^2}}{2}}, \tag{S43}$$

where

$$\psi = 2\frac{a+H}{c} - \xi_0\eta', \qquad \chi = \sqrt{(\xi_0^2-1)(1-\eta'^2)} \tag{S44}$$

By substituting $\rho_i$ in the form of Eq. (S27) and $\rho_0$ from Eq. (S38) to Eq. (S40) and integrating over $\phi'$, we obtain

$$\varphi_s(\eta) = -\beta \sum_{n=0}^{\infty} \sum_{m=0}^{\infty} (2n+1)\,(2m+1) v_n J_{nm} P_m(\eta) P_m(\xi_0) + \\ +\beta \frac{c(\varepsilon-1)E_0}{2\left(\varepsilon Q_1(\xi_0) - \xi_0 Q_0(\xi_0) + \frac{\xi_0^2}{\xi_0^2-1}\right)} \sum_{n=0}^{\infty} (2m+1)\, J_{1m} P_m(\eta) P_m(\xi_0), \tag{S45}$$

which can be written as

$$\varphi_i(\eta) = -\beta \sum_{n=0}^{\infty} \sum_{m=0}^{\infty} (2n+1)\,(2m+1) v_n J_{nm} P_m(\eta) P_m(\xi_0) + \\ +\beta \frac{cE_0}{2Q_1(\xi_0)} \sum_{n=0}^{\infty} (2m+1)\, J_{1m} P_m(\eta) P_m(\xi_0), \tag{S46}$$

where we have introduced the integral:

$$J_{nm} = \int_{-1}^{1} d\,\eta P_n(\eta) P_m(\eta_q) Q_m(\xi_q). \tag{S47}$$

Then, the continuity of the potential at the surface of the spheroid, Eq. (S20), can be written as follows:

$$2\sum_{m=0}^{\infty} (2m+1)\, v_m P_m(\xi_0) P_m(\eta) Q_m(\xi_0) + \varphi_s(\xi_0, \eta) = 2\sum_{m=0}^{\infty} (2m+1)\, U_m P_m(\xi_0) P_m(\eta), \tag{S48}$$

which brings us to the following equation:

$$2v_m Q_m(\xi_0) - \beta \sum_{n=0}^{\infty} (2n+1)\, v_n J_{nm} + \beta \frac{c(\varepsilon-1)E_0 J_{1m}}{2\left(\varepsilon Q_1(\xi_0) - \xi_0 Q_0(\xi_0) + \frac{\xi_0^2}{\xi_0^2-1}\right)} = 2U_m. \tag{S49}$$

The second boundary condition, Eq. (S21), brings us to the following system of equations:

$$2v_m P_m(\xi_0) \frac{\xi_0 Q_m(\xi_0) - Q_{m-1}(\xi_0)}{\xi_0 P_m(\xi_0) - P_{m-1}(\xi_0)} - \beta \sum_{n=1}^{\infty} (2n+1)\, v_n J_{nm} + \beta \frac{c(\varepsilon-1)E_0 J_{1m}}{2\left(\varepsilon Q_1(\xi_0) - \xi_0 Q_0(\xi_0) + \frac{\xi_0^2}{\xi_0^2-1}\right)} = 2\varepsilon U_m, \tag{S50}$$

Substituting $U_m$ from Eq. (S49) to Eq. (S50), we obtain

$$2v_m \left(\varepsilon \frac{Q_m(\xi_0)}{P_m(\xi_0)} - \frac{\xi_0 Q_m(\xi_0) - Q_{m-1}(\xi_0)}{\xi_0 P_m(\xi_0) - P_{m-1}(\xi_0)}\right) - \beta(\varepsilon-1) \sum_{n=1}^{\infty} (2n+1)\, v_n \frac{J_{nm}}{P_m(\xi_0)} = \\ = -\beta \frac{c(\varepsilon-1)^2 E_0 J_{1m}}{2P_m(\xi_0)\left(\varepsilon Q_1(\xi_0) - \xi_0 Q_0(\xi_0) + \frac{\xi_0^2}{\xi_0^2-1}\right)}. \tag{S51}$$

This can be written in simple notation:

$$\sum_{n=1}^{\infty} \mathcal{M}_{mn}\, v_n = \mathcal{C}_m. \tag{S52}$$

where

$$\mathcal{M}_{mn} = \delta_{nm} - \frac{\beta(\varepsilon-1)(2n+1)J_{nm}}{2P_m(\xi_0)\left(\varepsilon \frac{Q_m(\xi_0)}{P_m(\xi_0)} - \frac{\xi_0 Q_m(\xi_0) - Q_{m-1}(\xi_0)}{\xi_0 P_m(\xi_0) - P_{m-1}(\xi_0)}\right)}, \tag{S53}$$

$$\mathcal{C}_m = -\frac{\beta c(\varepsilon-1)^2 E_0 J_{1m}}{4P_m(\xi_0)\left(\varepsilon Q_1(\xi_0) - \xi_0 Q_0(\xi_0) + \frac{\xi_0^2}{\xi_0^2-1}\right)\left(\varepsilon \frac{Q_m(\xi_0)}{P_m(\xi_0)} - \frac{\xi_0 Q_m(\xi_0) - Q_{m-1}(\xi_0)}{\xi_0 P_m(\xi_0) - P_{m-1}(\xi_0)}\right)}. \tag{S54}$$

The system (S52) is infinite, however, it can be approximated by a finite system of equations, because for large $n$ and $m$, $\mathcal{C}_m$ tend to zero, and $\mathcal{M}_{mn}$ tend to the Kronecker symbol, $\delta_{nm}$. Indeed, $|P_n(x)| \leq 1$ if $x \in [-1,1]$ and $|Q_n(x)|$ monotonously tend to zero when $x$ tend to infinity [7]. Therefore, on the one hand, for large $m$, $\mathcal{C}_m$ behave as $\mathcal{C}_m \sim J_{1m}/Q_m(\xi_0)$, and $\mathcal{M}_m$ behave as $\mathcal{M}_m - \delta_{nm} \sim J_{nm}/Q_m(\xi_0)$. Om the other hand, $J_{nm}$ is the integral of $Q_m(\xi_q)$ multiplied by a limited function. Thus, since $\xi_q$ is always larger than $\xi_0$ (as it refers to the point outside of the spheroid), $J_{nm}/Q_m(\xi_0) \to 0$, when $m \to \infty$.

The final step in this section is to calculate the dipole moment of the prolate spheroid using the following expression:

$$p_z = c^2\xi_0^2 \int_{-1}^{1} \eta\, \rho(\eta) d\eta. \tag{S55}$$

Substituting $\rho_i$ from Eq. (S27) and $\rho_0$ from Eq. (S38) into Eq. (S55), we obtain the expression for the dipole moment of the prolate spheroid:

$$p_z = 2ac\upsilon_1 - \frac{(\varepsilon-1)c^3 E_0 P_1\left(\frac{a}{c}\right)}{3\left(\varepsilon Q_1\left(\frac{a}{c}\right) - P_1\left(\frac{a}{c}\right)Q_0\left(\frac{a}{c}\right) + \frac{a^2}{a^2-c^2}\right)}. \tag{S56}$$

Here, $\upsilon_1$ represents the solution to the system defined by Eq. (S52) for the index $n = 1$. It should be emphasized that the integral $J_{nm}$, as given by Eq. (S47), is independent on the sample's dielectric permittivity. As a result, all integrals $J_{nm}$ can be computed once and subsequently utilized across the entire frequency spectrum considered without significantly increasing the calculation time with the increase in the number of considered frequencies.

Significantly, $\upsilon_1$ exhibits direct proportionality to the external field $E_0$, due to its linear dependence on $C_m$, which itself scales proportionally with $E_0$. Consequently, the dipole moment component $p_z$ also demonstrates linear proportionality with $E_0$, enabling the introduction of polarizability, $\alpha_z = p_z/E_0$, defined as the ratio of the dipole moment to the external field, given explicitly by

$$\alpha_z = 2ac\vartheta_1 - \frac{(\varepsilon-1)c^3 P_1\left(\frac{a}{c}\right)}{3\left(\varepsilon Q_1\left(\frac{a}{c}\right) - P_1\left(\frac{a}{c}\right)Q_0\left(\frac{a}{c}\right) + \frac{a^2}{a^2-c^2}\right)}. \tag{S57}$$

where $\vartheta_n = \upsilon_n/E_0$.

## S4. DEPENDENCE OF THE SPECTRA ON TIP PARAMETERS

In this section, we validate our analytical solution by comparing it with numerical simulations performed using the finite-element method (FEM) implemented in the COMSOL Multiphysics software package. Subsequently, we compute near-field spectra for various tip parameters to examine how the spectra depend on the characteristics of the s-SNOM tip. Specifically, we consider a spheroid oriented vertically above the interface of a semi-infinite, homogeneous, and isotropic sample, placed in a uniform vertical electric field, $E_0 = 1$ V/m, as depicted in Fig. S3a. The system's behavior is governed by parameters such as the spheroid's length, $L$, its apex curvature radius, $R$, the separation distance between the spheroid and the sample surface, $H$, and the permittivities of both the spheroid, $\varepsilon_T$, and the sample material, $\varepsilon_S$. Given the electrostatic approximation, the frequency of the applied electric field does not explicitly appear in the governing equations but influences the permittivity of the involved materials. Here, we consider two example materials: gold, which behaves nearly as a perfect electric conductor (PEC) within the infrared (IR) spectral range, and SiC, known for its pronounced near-field response [8]. For SiC, we adopt the following analytical model to describe its permittivity [8]:

$$\varepsilon_{SiC}(\omega) = \varepsilon_\infty \left(1 + \frac{\omega_{LO}^2 - \omega_{TO}^2}{\omega_{TO}^2 - \omega^2 - i\gamma_l\omega} - \frac{\omega_p^2}{\omega^2 + i\gamma_p\omega}\right) \tag{S58}$$

where $\varepsilon_\infty = 6.56$, $\omega_{LO} = 971\ \text{cm}^{-1}$, $\omega_{TO} = 797\ \text{cm}^{-1}$, $\gamma_l = 3.3\ \text{cm}^{-1}$, $\omega_p = 275\ \text{cm}^{-1}$ $\gamma_p = 450\ \text{cm}^{-1}$, as shown in Fig. S3b.

## A. Spheroid dipole moment as a function of the tip-sample distance

Figures S3c and S3d illustrate the amplitude and phase of the spheroid's dipole moment at $\omega = 940\ \mathrm{cm}^{-1}$ as functions of the separation distance between the spheroid and the sample surface, $H$. For gold, the amplitude of the dipole moment monotonically decreases with increasing distance, while the phase remains close to zero, reflecting the nearly PEC behavior of both the spheroid and the sample materials. Conversely, in the case of SiC, the amplitude and phase exhibit non-monotonic behavior with distinct peaks indicating resonant interactions between the spheroid and the sample.

Figures S3e and S3f depict the spheroid dipole moment over the SiC surface as a function of frequency for several fixed $H$ values. At shorter distances, multiple resonant peaks in amplitude become evident. The excellent consistency between analytical and numerical results across the entire frequency and distance range validates the accuracy of our analytical solution.

To elucidate the origin of multiple resonances, we analyze the field distributions at three representative spheroid-sample distances corresponding to two maxima and one intermediate minimum in the dipole moment amplitude shown in Fig. S3c. As shown in Figure S3g, distinct maxima correlate with different orientations of the local electric field between the spheroid and the sample surface. At larger distances, the local field aligns with the external field, whereas at shorter distances, the field orientation becomes reversed. Consequently, the multiple peaks observed in the dipole moment as a function of $H$ correspond to different resonant modes within the spheroid-sample system when the sample permittivity ranges approximately between $-1$ and $-5$.

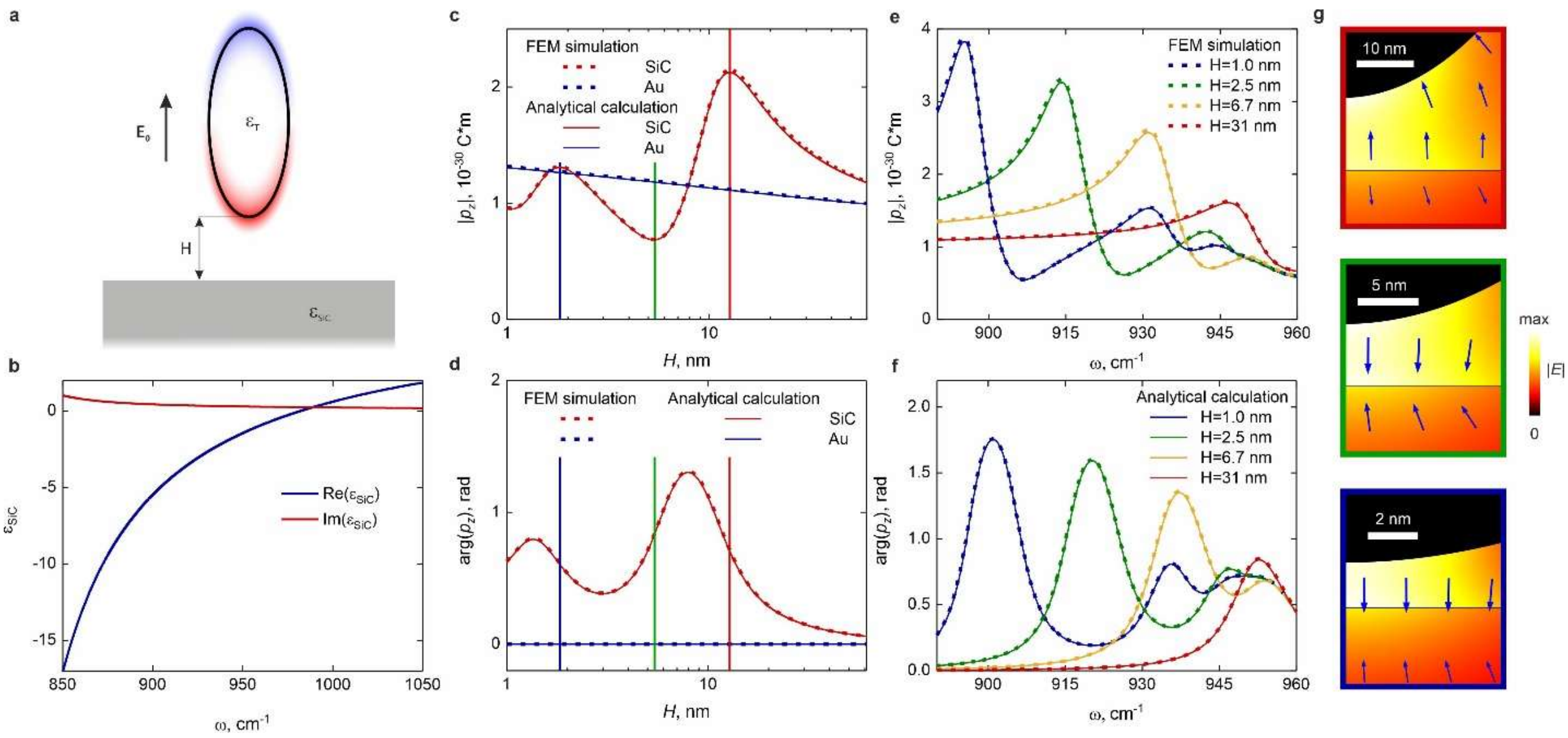

FIG. S3. a) Schematic of the spheroid over the sample surface in a uniform electric field, $\mathbf{E}_0$. b) The dielectric permittivity of the SiC sample. c, d) Amplitude and phase of the dipole moment of the spheroid placed over the SiC and gold surfaces as a function of the distance between the spheroid apex and the sample surface at the frequency $\omega = 940\ \mathrm{cm}^{-1}$. e, f) Amplitude and phase of the dipole moment of the spheroid placed at different distances over the SiC surface as a function of the frequency. g) Distributions of the magnitude of the electric field near the apex of the spheroid placed over the SiC surface at the frequency $\omega = 940\ \mathrm{cm}^{-1}$. The arrows show the directions of the electric field. The frame color corresponds to the height marked with the same colors in panels c and d. The geometrical parameters of the spheroid are $R = 25$ nm, $L = 600$ nm. The external field $E_0 = 1$ V/m.

## B. Variation of the model parameters

Next, we compute the demodulated dipole moment of the spheroid over SiC, normalized by the demodulated dipole moment of the spheroid over gold, following the standard procedure used in experimental data processing. Figure S4 illustrates the influence of various modulation parameters on the resulting spectra, namely, the spheroid oscillation amplitude, $A$, the minimum distance between the spheroid and the sample surface, $H_0$, and the harmonic number of the demodulated signal, $n$. In these calculations, all parameters except the sample permittivity are identical for gold and SiC. As clearly shown in the figure, the spectral characteristics are significantly influenced by these modulation parameters. While variations in the harmonic number and oscillation amplitude primarily affect the amplitude of the spectral peaks, changes in $H_0$ also markedly alter the shape and position of these peaks.

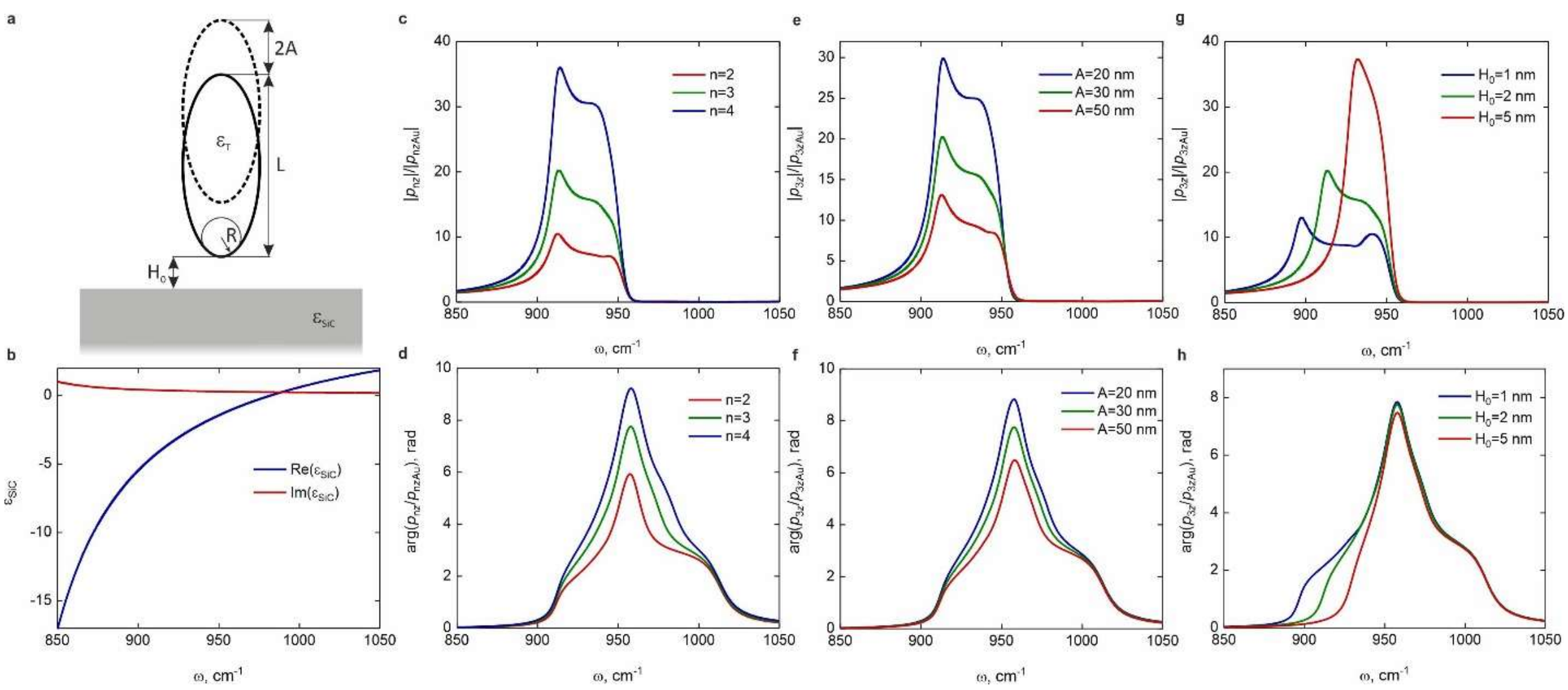

FIG. S4. a) Schematic of the oscillating spheroid with its geometrical parameters. b) The dielectric permittivity of the SiC sample. c-h) Amplitude (c,e,g) and phase (d,f,h) spectra of the demodulated dipole moment of the spheroid over the sample normalized to those of the spheroid over the gold surface for different demodulation harmonics (c,d), oscillation amplitudes (e,f), and the minimal distances between the tip apex and sample surfaces (g,h). In panels c and d, $A = 30$ nm, $H_0 = 2$ nm; in panels e and f, $H_0 = 2$ nm, $n = 3$; in panels g and h, $A = 30$ nm, $n = 3$. The geometrical parameters of the spheroid are $R = 25$ nm, $L = 600$ nm; the dielectric permittivity of the spheroid is $\varepsilon_T = -10^{10}$.

Next, we investigate how the spheroid parameters influence the near-field spectrum. Figures S5a and S5b present the spectra obtained for spheroids with varying dielectric permittivities, $\varepsilon_T$. A notable change in both the spectral peak shape and amplitude occurs primarily when transitioning from a metallic ($\varepsilon_T < 0$) to a dielectric spheroid ($\varepsilon_T = 4$). In contrast, the spectral differences between a poorly conducting metal ($\varepsilon_T = -100 + 50i$) and PEC ($\varepsilon_T = -10^{10}$) are relatively minor. Figures S5c and S5d illustrate that the radius of curvature, $R$, significantly affects both the peak shape and its position. The spheroid's length, $L$, however, mainly impacts the amplitude of the spectral peak and is found to be the least influential parameter: a 2.5-fold variation in $L$ results in only about a 20% change in peak amplitude.

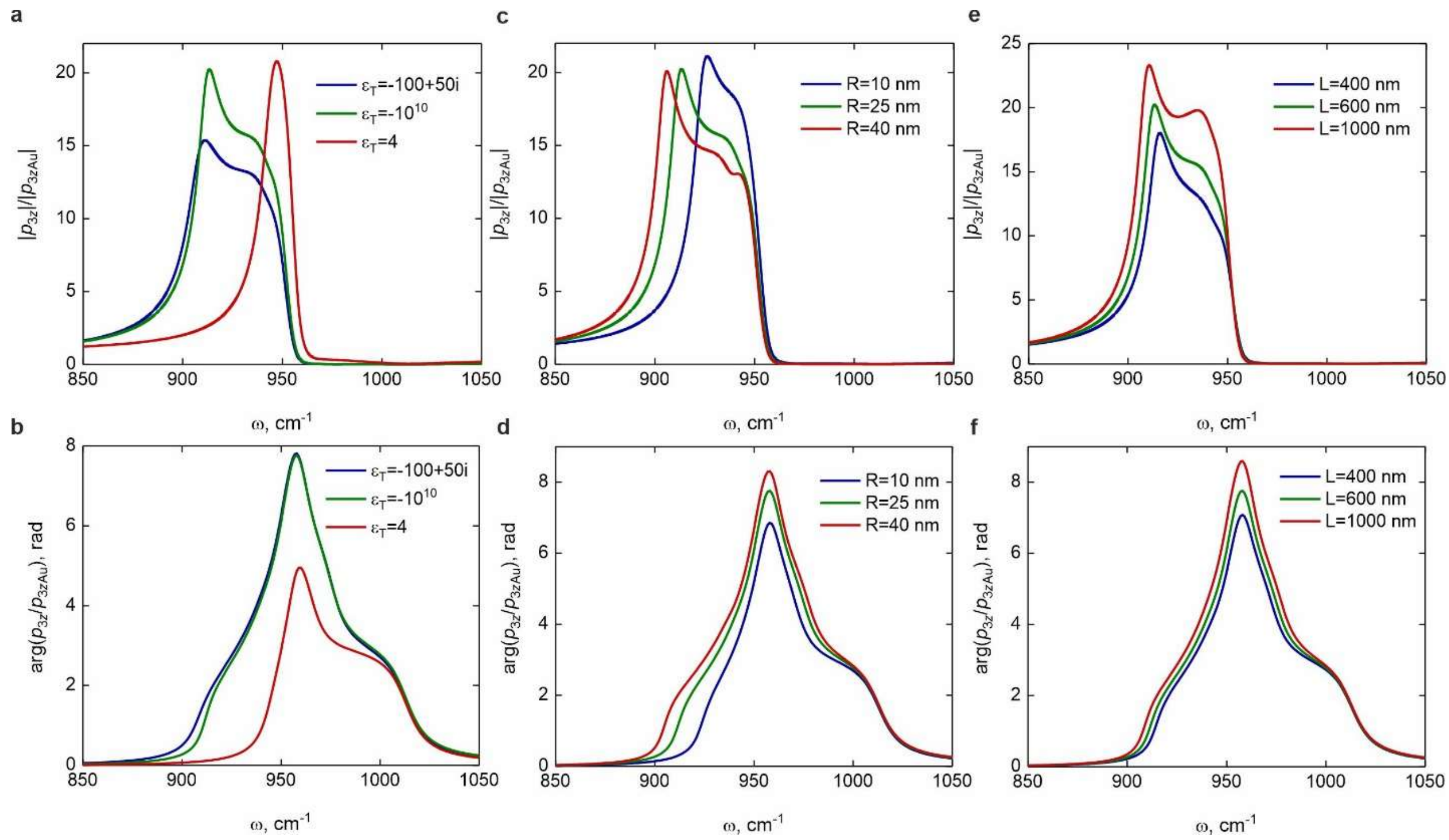

FIG. S5. Third harmonic amplitude (a,c,e) and phase (b,d,f) spectra of the dipole moment of the spheroid over the sample normalized to those of the spheroid over the gold surface for different dielectric permittivities of the spheroid (a,b), spheroid apex curvature radiuses (c,d), and the spheroid lengths (e,f). In panels a and b, $R = 25$ nm, $L = 600$ nm; in panels c and d, $\varepsilon_T = -10^{10}$, $L = 600$ nm; in panels e and f, $R = 25$ nm, $\varepsilon_T = -10^{10}$. The oscillation parameters of the spheroid are $A = 30$ nm, $H_0 = 2$ nm.

## C. Approach curves

The results presented in Figs. S4 and S5 highlight the particular significance of the parameter $H_0$. Firstly, its effect on the spectra is highly pronounced; yet, it is currently among the least controlled and least accurately measurable parameters, primarily due to significant uncertainties introduced by interaction forces between the tip and the surface. Secondly, unlike other parameters that can be assumed identical for both the sample and the reference material—even if their precise values are unknown—the parameter $H_0$ is inherently dependent on the specific properties of the sample surface, making it impossible to guarantee its consistency across different materials. Consequently, a detailed theoretical and experimental investigation of the influence of $H_0$ represents a critical area for future research.

To establish a foundation for such studies, Fig. S6 presents the so-called approach curves, illustrating the dependence of the demodulated spheroid dipole moment on the minimum distance between the spheroid and the sample surface at a frequency of $\omega = 940$ cm$^{-1}$. For gold, the amplitude of the dipole moment exhibits a monotonic decrease, as anticipated. Conversely, the behavior for SiC is more complex, consistent with expectations derived from Figure S3. Specifically, the demodulated dipole moment decreases notably only after $H_0$ surpasses the position of the last and most pronounced peak depicted in Fig. S3c. The intermediate peaks observed previously are effectively smoothed out due to spheroid oscillations, resulting in a flattened, plateau-like shape in the approach curve at smaller values of $H_0$. Therefore, $H_0$ between the spheroid and the dispersive sample predominantly influences the shape and position of the observed peak, whereas $H_0$ between the spheroid and the reference material significantly affects the amplitude of the spectrum after normalization.

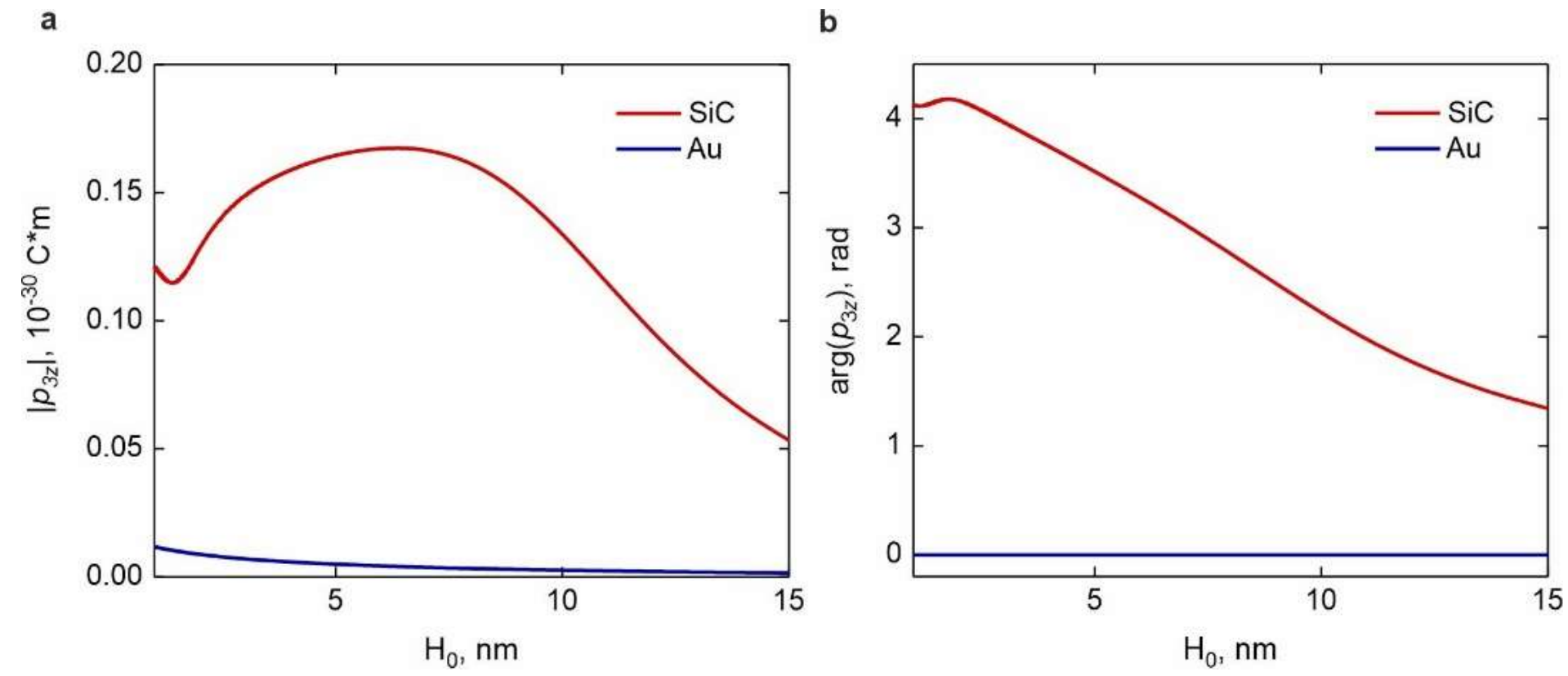

FIG. S6. Third harmonic amplitude (a) and phase (b) of the spheroid over the sample and gold surfaces at frequency $\omega = 940\ \text{cm}^{-1}$ as a function of the minimal distance between the spheroid and the surface. The geometrical parameters of the spheroid are $R = 25$ nm, $L = 600$ nm; the dielectric permittivity of the spheroid is $\varepsilon_T = -10^{10}$; the oscillation amplitude is $A = 30$ nm.

## S5. EXPERIMENTAL METHODS

We used a commercial nano-FTIR setup (Attocube Systems AG, Germany), in which the oscillating (at a frequency $\Omega = 252$ kHz; tapping amplitude of $40$ nm) Pt/Ir-coated AFM tip (Arrow-NCPt-50, NanoWorld AG, Switzerland) was illuminated by p-polarized mid-IR broadband radiation generated by a supercontinuum laser (Femtofiber pro IR and SCIR; Toptica, Gräfelfing, Germany; average power of $0.5$ mW; frequency range $800 - 1400\ \text{cm}^{-1}$). The detector signal was demodulated at a frequency $n\Omega$ (with $n = 3, 4$) for effective background suppression. Interferograms were measured by recording the demodulated detector signal as a function of the position of the reference mirror, $d$, at a fixed tip position. For apodization of the interferograms, a Tukey window function with $\alpha = 0.3$ was applied. After zero-filling (3× padding) we Fourier transformed the interferograms to obtain complex-valued near-field point spectra, $E_s(\omega)$. Each spectrum is an average of 3 spectra recorded. We normalized the obtained spectra to a reference spectrum recorded on gold, $E_{s,Au}(\omega)$, measured under the same conditions and parameters. The number of pixels for each interferogram was 4096, with an integration time of $15$ ms. Thus, each measurement lasted 3 minutes. The spectral resolution was $6.25\ \text{cm}^{-1}$, the limit of the setup.

## S6. PUBLICLY AVAILABLE NUMERICAL TOOL

To support further research and encourage broader adoption of our method, we have implemented the model in a publicly available numerical tool: a standalone MATLAB-based application. This application does not require a full MATLAB installation; it only relies on the free MATLAB Runtime library, which is automatically downloaded during installation if not already present. The tool enables users to compute spectra of bulk samples with customizable tip and sample parameters.

Additionally, we provide the MATLAB script used for the calculations presented in Section 4. The script includes detailed comments to aid understanding and facilitate modification, making it easier to adapt the method or its components for calculations beyond the scope of the standalone application.